\documentclass[letterpaper,twocolumn,10pt]{article}
\usepackage{usenix2019_v3}

\usepackage{cite}
\usepackage{algorithmic}
\usepackage{textcomp}
\usepackage{booktabs}
\usepackage{multirow}
\usepackage{graphicx}
\usepackage{xspace}
\usepackage{pifont}
\usepackage{amsmath,amssymb,amsfonts}
\usepackage{listings}
\usepackage{xcolor}
\usepackage{tikz}
\usepackage{import}
\usepackage[normalem]{ulem}
\usepackage{makecell}
\usepackage{mathtools}
\usepackage{framed}
\usepackage{tabularx}
\usepackage{subfig}
\usepackage{cite}
\usepackage{amsmath,amssymb,amsfonts}
\usepackage{algorithmic}
\usepackage{graphicx}
\usepackage{textcomp}
\usepackage{xcolor}
\usepackage{amsthm}

\def\isindraft{0}
\newcommand{\indraft}{\def\isindraft{1}}
\renewcommand{\paragraph}{\vspace{2pt}\noindent \textbf}

\usepackage{etoolbox}

\newcommand{\shmem}{ShMem\xspace}

\newcommand{\create}{{$\tt{create}$}\xspace}
\newcommand{\map}{{$\tt{map}$}\xspace}
\newcommand{\unmap}{{$\tt{unmap}$}\xspace}
\newcommand{\share}{{$\tt{share}$}\xspace}
\newcommand{\change}{{$\tt{change}$}\xspace}
\newcommand{\destroy}{{$\tt{destroy}$}\xspace}
\newcommand{\transfer}{{$\tt{transfer}$}\xspace}

\newcommand{\uid}{{$\tt{uid}$}\xspace}

\newcommand{\futex}{{$\tt{futex}$}\xspace}

\newcommand{\stname}[1]{\emph{#1}}

\newcommand{\stspatial}{\stname{spatial}\xspace}
\newcommand{\stcodename}{\stname{\codename}\xspace}

\newcommand{\stnative}{\stname{native}\xspace}

\newcommand{\codename}{\textsc{Elasticlave}\xspace}

\newcommand{\riscv}{\mbox{RISC-V}\xspace}

\providecommand{\loc}{LoC\xspace}

\definecolor{lightgray}{rgb}{.9,.9,.9}
\definecolor{darkgray}{rgb}{.4,.4,.4}
\definecolor{purple}{rgb}{0.65, 0.12, 0.82}

\newcommand{\mPrincipal}{\mathtt{EnclaveID}}
\newcommand{\mRegion}{\mathtt{UID}}
\newcommand{\mPermission}{\mathtt{Permission}}
\newcommand{\mVaddr}{\mathtt{Vaddr}}
\newcommand{\mSize}{\mathtt{Size}}
\newcommand{\mByte}{\mathtt{Byte}}
\newcommand{\mOffset}{\mathtt{Offset}}
\newcommand{\mP}{\mathcal{P}}
\newcommand{\mR}{\mathcal{R}}
\newcommand{\mA}{\mathcal{A}}
\newcommand{\mM}{\mathcal{M}}
\newcommand{\mV}{\mathcal{V}}
\newcommand{\mS}{\mathsf{S}}
\newcommand{\mSp}{\mathsf{S}^\prime}
\newcommand{\mError}{\mathtt{Error}}
\newcommand{\mSuccess}{\mathtt{SUCCESS}}
\newcommand{\dom}{\mathrm{dom}}
\newcommand{\mintersect}{\mathrm{Intersect}}
\newcommand{\mcovers}{\mathrm{Covers}}

\newcommand{\targetfreq}{$800$~MHz}

\begin{document}

\title{\codename: An Efficient Memory Model for Enclaves}

\date{}

\author{
{\rm Zhijingcheng Yu}\\
{\normalsize National University of Singapore}
\and
{\rm Shweta Shinde}~\thanks{Part of the research was done while at University of California, Berkeley.}\\
{\normalsize ETH Zurich}
\and 
{\rm Trevor E. Carlson}\\
{\normalsize National University of Singapore}
\and
{\rm Prateek Saxena}\\
{\normalsize National University of Singapore}
}  

\maketitle

\thispagestyle{empty}

\indraft

\begin{abstract}
Trusted-execution environments (TEE), like Intel SGX, isolate
user-space applications into secure enclaves without trusting the OS.
Thus, TEEs reduce the trusted computing base, but add one to two orders of
magnitude slow-down. The performance cost stems from a strict memory
model, which we call the {\em spatial isolation model}, where enclaves cannot share memory regions with each other.
In this work, we present \codename---a new TEE memory model that
allows enclaves to selectively and temporarily share memory with other
enclaves and the OS. 
\codename eliminates the need for expensive data copy operations, while
offering the same level of application-desired security as possible
with the spatial model.
We prototype \codename design on an RTL-designed cycle-level RISC-V core
and observe $1$ to $2$ orders of magnitude performance improvements over
the spatial model implemented with the same processor configuration. \codename has a small TCB. We find that its performance characteristics and hardware area footprint scale well with the number of shared memory regions it is configured to support.
\end{abstract} \section{Introduction}

Isolation, commonly using OS processes, is a cornerstone abstraction
for security. It allows us to isolate and limit software compromises
to one {\em fault domain} within an application and is the basis for
implementing the design principle of privilege separation. In the last
few years, user-level enclaves have become available in commodity CPUs
that support TEEs. A prime example of enclaved TEEs is Intel
SGX~\cite{sgxexplained}. Conceptually, enclaves are in sharp contrast
to processes in that they do not trust the privileged OS, promising a
drastic reduction in the TCB of a fault domain. This is why the design
of enclaved TEEs is of fundamental importance to security.

One of the big challenges with using today's enclaves is {\em
performance}. For example, many prior efforts have reported $1$--$2$
orders of magnitude slowdowns when supporting common
applications on SGX~\cite{ratel,graphene-sgx, scone, haven,occlum}. 
This raises the question whether one can design enclaved TEEs which
have substantially better performance. 

As a step towards this goal, we point towards one of the key
abstractions provided by enclaved TEEs---their memory model. The memory
model used in several existing TEEs
~\cite{armtrustzone,amd-sev,ferraiuolo2017komodo,
sgx,mckeen2016sgx,costan2016sanctum}, including SGX,
which originates from a long line of prior
works~\cite{bastion,overshadow}, follows what we call the {\em spatial
isolation model}. In this model, the virtual memory of the enclave is
statically divided into two types: {\em public} and {\em private}  
memory 
regions. These types are fixed throughout the region's lifetime. The
spatial isolation model is a simple but a rigid model, as its
underlying principle breaks compatibility with the most basic of data
processing patterns where data needs to privately computed on before
being made public or shared externally. In the spatial model, 
 traditional applications will need 
to create multiple data copies when sharing across enclave
boundaries, and additionally encrypt data, if they desire security
from an untrusted OS. Consequently, to support abstractions like
shared memory, pipes, fast synchronization, IPC, file I/O, and others
on spatially-isolated memory, data has to be copied between public to
private memory regions frequently. This results in very high
 overheads, a phenomenon reported in many frameworks
trying to re-enable compatibility on TEEs that use the spatial
model~\cite{ratel,occlum,graphene-sgx,scone,haven,dayeol2020keystone}.

In this work, we revisit the spatial isolation memory model adopted by
modern TEEs. We propose a new memory model called \codename which
allows enclaves to share memory across enclaves and with the OS, with
more flexible permissions than in spatial isolation.
While allowing flexibility, \codename does not make any simplistic
security assumptions or degrade its security guarantees over the
spatial isolation model. We view enclaves as a fundamental
abstraction for partitioning applications in this work, and therefore,
we assume that enclaves do {\em not} trust each other and can become 
compromised during their lifetime.
The \codename design directly eliminates the need for expensive data
copy operations, which are necessary in the spatial isolation model to
ensure security. The end result is that \codename offers
$10\times$--$100\times$ better performance than spatially-isolated
TEEs with the same level of application-desired security.

The main challenge designing \codename is providing sufficient
flexibility in defining security over shared memory regions, while
{\em minimizing complexity}. Relaxing the spatial isolation model such
that it allows enclaves to privately share memory between them,
without trusting a trusted OS as an intermediary, requires careful
consideration. 
In particular, we want to allow enclaves to share a memory region and
be able to alter their permissions on the region over time, thereby
eliminating the need to create private copies. The permission
specification mechanism should be flexible enough to allow non-faulty
(uncompromised) enclaves to enforce any desirable sequence of
permission changes on the region which the application demands. At the
same time, we do {\em not} want the compromised OS or any other
enclaves that may have become compromised during runtime to be able to
escalate their privileges arbitrarily, beyond what was initially
agreed  upon.
For instance, simply providing the equivalent of the traditional
shared memory and IPC interfaces (e.g., POSIX) can leave several
avenues of attacks unaddressed. The untrusted OS or
compromised enclaves may modify/read shared memory out of turn, create
TOCTOU attacks, intentionally create race conditions, re-delegate
permissions, and so on.
Thus, the \codename interface is designed with abstractions that relax
the spatial model {\em minimally}. Further, a simple interface design
makes it easy to analyze the final security, and simultaneously, keeps
the implementation impact small.

We implement our design on an open-source, RTL-level \riscv
\targetfreq{}
processor~\cite{rocket}. 
We evaluate performance and chip area impact of \codename using a 
cycle-level simulator~\cite{DBLP:conf/isca/KarandikarMKBAL18} on
several synthetic as well as real-world benchmarks. We observe that \codename enables performance improvements of {\em $1$--$2$ orders of magnitude} over the spatial isolation model implemented in the same
processor configuration. We also show that \codename has a modest cost
on implementation complexity. First, we show that the additional TCB
is than $7,000$ LoC. 
Second, our benchmarking highlights that the performance overhead is
affected primarily by the number of enclave-to-enclave context
switches, i.e, it is independent of the size of shared data in a
region. Further, the increased hardware register pressure due to
\codename does {\em not} increase the critical path of the synthesized
core for all tested configurations. 
Third, the hardware area footprint scales well with the maximum number
of shared regions \codename is configured to support. Specifically,
our estimated hardware area increase is below $1\%$ of our baseline
\riscv processor, for every $8$ additional shared memory regions
\codename TEE is configured to support. 

\paragraph{Contributions.}
The paper proposes a new memory model for enclaved TEEs called
\codename. We show that its design can result in significantly better
performance than the spatial isolation model. We offer a prototype
implementation on a \riscv processor, with a modest hardware area impact.
 \section{Problem}
\label{sec:problem}
TEE provides the abstraction of enclaves to isolate components of an application, which run with user-level privileges.
The TEE implementation (privileged hardware) is trusted and assumed to be bug-free.\footnote{TEEs are typically implemented in hardware and firmware. Our TEE  implementation uses \riscv hardware feature along with a privileged software monitor, executing in the firmware-equivalent software privileged layer.}
We want to design an efficient memory model for TEEs that support enclaves.
In our setup, a security-sensitive application is partitioned into multiple potentially
compromised (or faulty)  components. Each component runs in a
separate enclave created by the TEE, which serves as a basic isolation primitive. 
Enclaves are assumed to be {\em mutually-distrusting}, since they can
be compromised by the adversary during their execution, e.g., due to
software exploits. This assumption is of fundamental importance, as it
captures the essence of why the application is partitioned to begin
with. The memory model adopted by Intel SGX serves as a tangible
baseline to explain how a rigid security models can induce prohibitive
performance costs.

\begin{figure*}[!ht]
    \centering
    \subfloat[Producer-consumer\label{fig:problems-prod-cons}]{
        \raisebox{-0.5\height}{\includegraphics[scale=0.33]{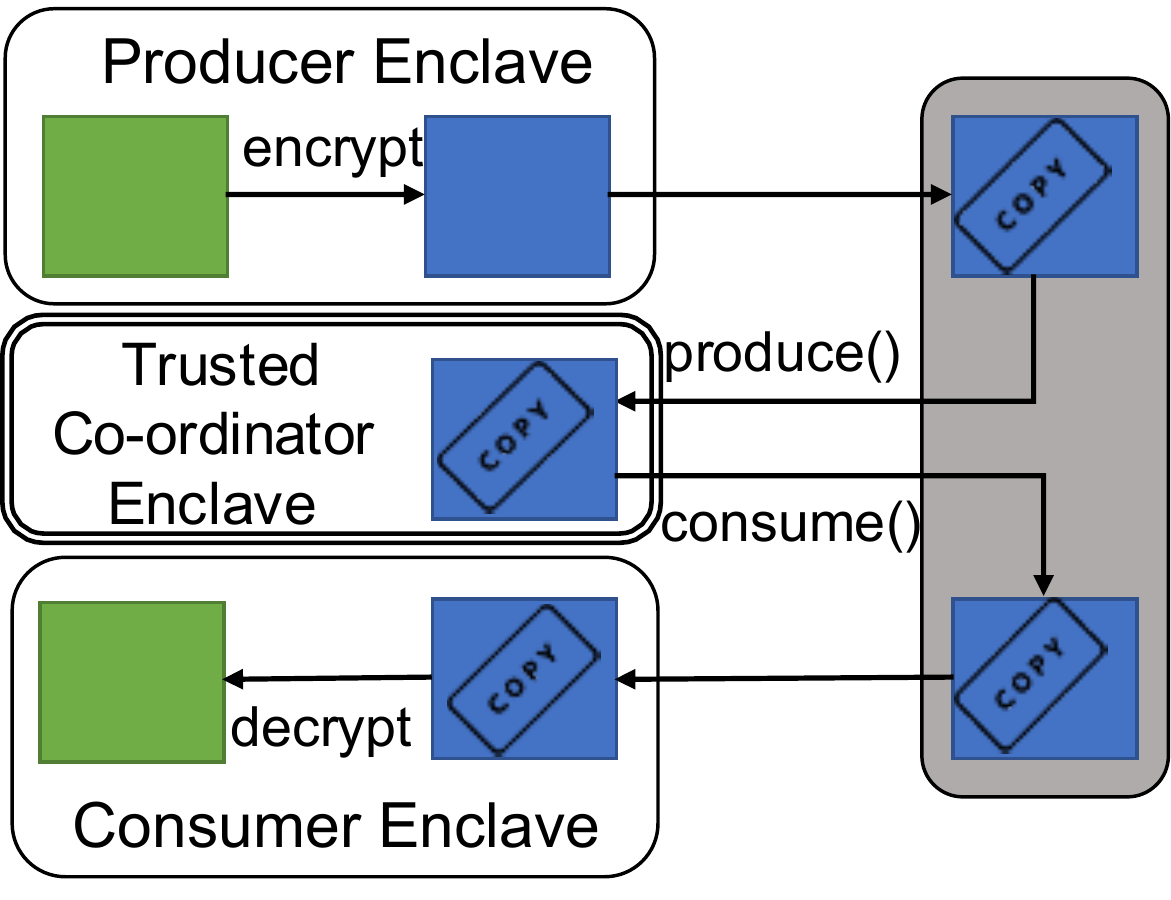}}
    }
    \hspace*{0.05\linewidth}
    \subfloat[Proxy\label{fig:problems-proxy}]{
        \raisebox{-0.5\height}{\includegraphics[scale=0.35]{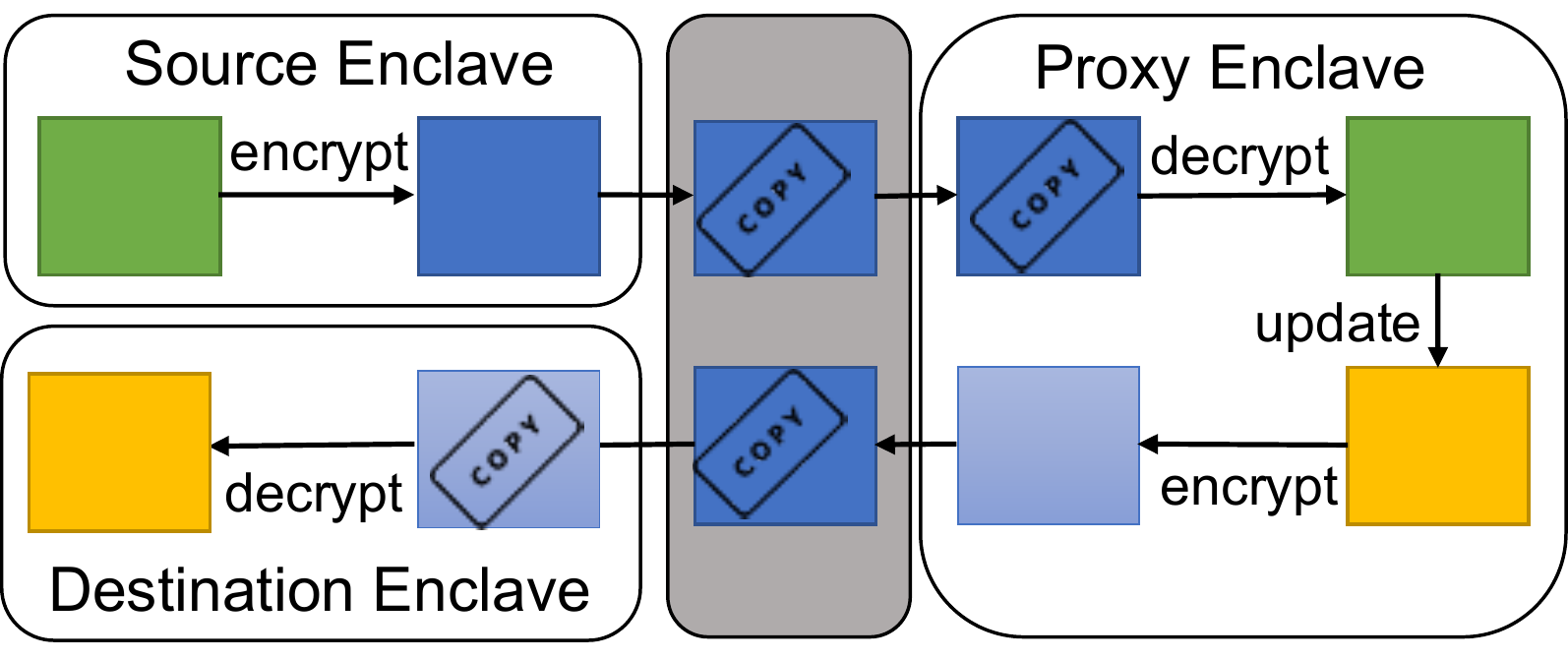}}
    }
    \hspace*{0.05\linewidth}
    \subfloat[Client-server\label{fig:problems-client-server}]{
        \raisebox{-0.5\height}{\includegraphics[scale=0.35]{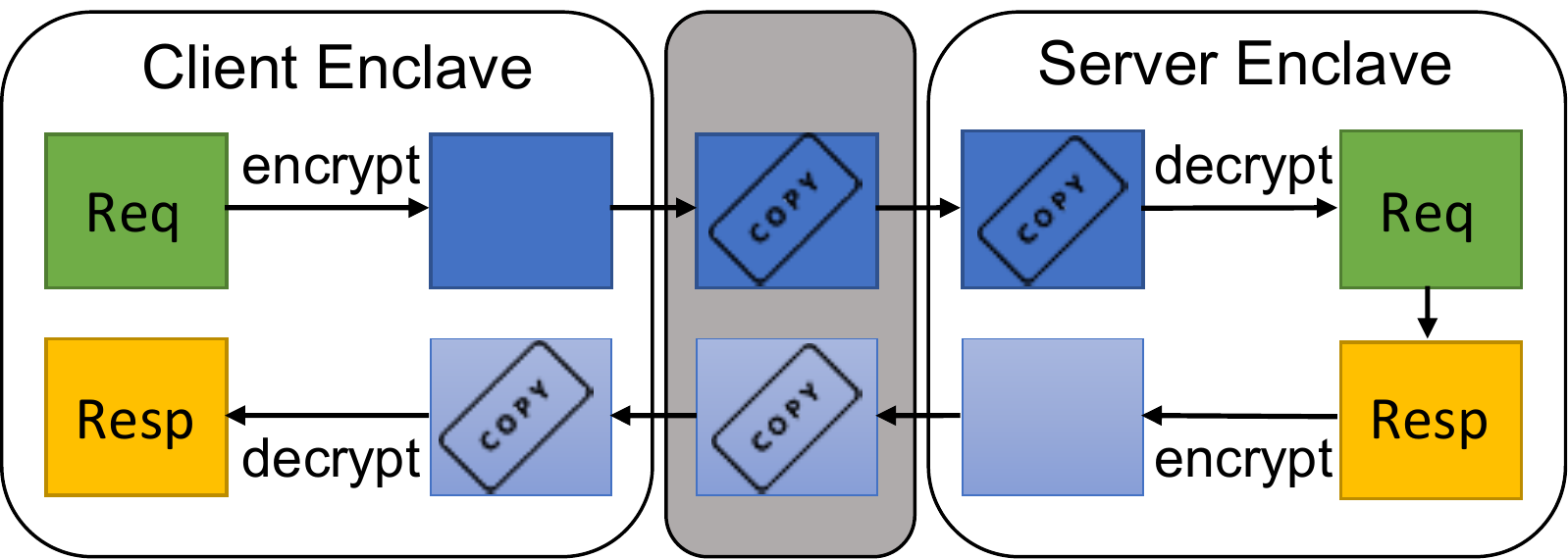}}
    }
    \caption{Spatial \shmem baseline cost.
    Gray color indicates public memory; 
    double-line border indicates the trusted coordinator.}
    \label{fig:problems}
\end{figure*}
 
\subsection{Baseline: The Spatial Isolation Model}
\label{sec:baseline-model}

Most enclaves available on commodity processors, use a memory model
which we call the \emph{spatial isolation}
model~\cite{sgx,sgx2014progref, bastion, xom, aegis, mi6}, including Intel SGX, which follows many prior 
proposals~\cite{bastion,overshadow}. In this model, each enclave 
comprises two different types of non-overlapping virtual memory regions: 
\begin{enumerate}
\item \emph{Private memory:} 
exclusive to the enclave itself and inaccessible to all other
enclaves running on the system.
\item \emph{Public memory:} 
fully accessible to the enclave and the untrusted OS, who  may then 
share it with other enclaves.
\end{enumerate}

The spatial model embodies the principle of dividing trust in an ``all
or none'' manner~\cite{ratel}. For each enclave, every other enclave
is fully trusted to access the public memory, whereas the private
memory is accessible only to the enclave itself. 
This principle is in sharp contrast to any form of memory sharing,
which is extensively used when an enclave wants to exchange data with
the outside world, including with other enclaves. Memory sharing is
key to efficiency in I/O operations, inter-process communication,
multi-threading, memory mapping, signal-handling, and other standard
abstractions. Although shared memory is not directly possible to
implement in the spatial isolation model, it can be {\em simulated}
with message-passing abstractions instead. To discuss the limitations
of the spatial isolation concretely, we present a baseline for
implementing shared memory functionality in the spatial model next. 
We refer to this baseline as the {\em spatial \shmem baseline.} Note
that this baseline is frequently utilized in many prior frameworks
that offer compatibility with Intel SGX~\cite{ratel, scone, graphene-sgx}.

\paragraph{The Spatial \shmem Baseline.}
This baseline simulates a shared memory abstraction between two
spatially isolated enclaves.
Observe that the two enclaves can keep a private copy of the shared
data. However, as the enclaves do not trust each other they cannot 
access each other's local copy. Therefore, the shared data must
either reside in public memory, which may be mapped in the address
space of both the enclaves, or they must use message-passing (e.g., via
RPC) which itself must use the public memory. Any data or exchanged
messages using public memory are exposed to the untrusted OS.
Therefore, the spatial \shmem baseline requires a cryptographically  secure
channel to be implemented on top of the public memory. Specifically,
the two enclaves encrypt data and messages before writing them to
 public memory and decrypt them after reading them. We call this
a {\em secure public memory}. We can assume that the cryptographic
keys are pre-exchanged or pre-established securely by the enclaves.

A secure public memory is not sufficient to implement a shared memory
abstraction in the spatial \shmem baseline. Concurrently executing
enclaves may access data simultaneously and such accesses may require serialization in order to 
maintain typical application consistency guarantees.
Notice that reads and writes to the secure public channel involves
encryption and decryption sub-steps, the atomicity of which is not
guaranteed by the TEE. No standard synchronization primitives
such as semaphores, counters, and futexes---which often rely on
OS-provided atomicity---remain trustworthy in the enclave threat model
we consider. Therefore, one simple way to serialize access is to use a 
trusted mediator or coordinator enclave.
In the spatial \shmem baseline, we designate a third enclave as a
{\em trusted coordinator}. For achieving memory consistency, accesses
to shared memory are simulated with message-passing, i.e., read/writes
to shared memory are simulated with remote procedure calls to the
trusted coordinator, which implements the ``shared'' memory by
keeping its content in its private memory. For example, to implement a
shared counter, the trusted coordinator enclave keeps the counter in
its private memory, and the counter-party enclave can send messages to
the trusted coordinator for state updates. 

We have assumed in our baseline that the coordinator is not faulty or
compromised. Attacks on the trusted coordinator can subvert the
semantic correctness of the shared memory abstraction. One can
consider augmenting this baseline to tolerate faulty coordinators
(e.g., using BFT-based mechanisms). But these mechanisms would only {\em increase} the performance costs and latencies, 
reducing the overall throughput.

\subsection{Illustrative Performance Costs}
\label{sec:pattern-costs}

The spatial \shmem baseline is significantly more expensive to
implement than the original shared memory abstraction in a non-enclave
(native) setting. We refer readers to Section~\ref{sec:eval} for the
raw performance costs of the spatial \shmem baseline over the native.
The overheads can be $1$-$2$ orders of magnitude higher. This is
primarily because of the encryption-decryption steps and additional memory
copies that are inherent in the implementation of secure channel and
trusted coordinator. Several recent works have reported these costs
over hundreds of programs on the Intel SGX platform~\cite{occlum,
graphene-sgx, scone, ratel}. For instance Occlum reported overheads up
to $14\times$ as compared to native Linux execution~\cite{occlum}. We
present $3$ representative textbook patterns of data sharing that ubiquitously arise
in real-world applications and illustrate how spatial isolation incurs
such significant cost.

\paragraph{Pattern 1: Producer-Consumer.}
In this pattern, the producer enclave writes a stream of
data objects to shared memory for a consumer enclave to read and
process. Several applications use this for signaling completion of
sub-steps of a larger task, such as in MapReduce~\cite{vc3, m2r}. For
supporting this pattern with the spatial \shmem baseline, the producer
has to copy its output data to public memory first and then the
consumer enclave copies it to private memory. In summary, at least $2$
additional copies of the original shared data are created. Further,
the data is encrypted and decrypted once leading to $2$ compute
operations per byte and $1$ private copy in the trusted coordinator.
Figure~\ref{fig:problems-prod-cons} depicts the steps.

\paragraph{Pattern 2: Proxy.}
Many applications serve as a intermediate proxy between a producer and
consumer. For example, consider packet monitoring/filtering
application like \texttt{Bro}, \texttt{snort}, or \texttt{bpf} which
modifies the data shared between two end-point
applications. Proxy designs can be implemented using two instances of
the producer-consumer pattern, where the proxy acts as the consumer
for the first instance and producer for the second. However, in
practice, proxies often optimize by performing {\em in-place} changes
to the shared data rather than maintaining separate queues with the
end points~\cite{vif, sgx-box}. Such in-place memory processing is not
compatible with the spatial memory model. Applications which
originally use this pattern must incur additional memory copies. The
data stream must be placed in public memory, so that it can be passed to 
the proxy enclave that acts as a trusted coordinator. But at the same
time, the proxy cannot operate on public memory in-place, or else it
would risk modifications by other enclaves. Therefore, there are at
least $2$ memory copies of the $2$ original shared data contents,
totaling $4$ copies when supporting this pattern with the spatial
\shmem baseline, as shown in Figure~\ref{fig:problems-proxy}. Further,
the data is encrypted and decrypted twice leading to $4$ compute
operations per byte.

\paragraph{Pattern 3: Client-Server.}
Two enclaves, referred to a client and a server, read and write shared
data to each other in this pattern. Each enclave reads the data
written by the other, performs private computation on it, and writes
back the computed result back to the shared memory. As explained, the
shared memory abstraction cannot directly be implemented with data
residing in a shared region of public memory since the OS and other
enclaves on the system are not trusted. For supporting such sharing
patterns, there will be at least $4$ data copies---one in server
private memory, one client private memory, and two for passing data
between them via a public memory. Further, the data is encrypted and
decrypted twice leading to $4$ compute operations per byte
(Figure~\ref{fig:problems-client-server}).

\begin{table}[!t]
\centering
\resizebox{0.4\textwidth}{!}{
\begin{tabular}{@{}lllllc@{}}
\toprule
\multicolumn{2}{c}{\multirow{2}{*}{\textbf{Pattern}}} & \multicolumn{3}{c}{\textbf{\begin{tabular}[c]{@{}c@{}}Spatial\end{tabular}}} & \textbf{\codename} \\ \cmidrule(l){3-6} 
\multicolumn{2}{c}{} & \textbf{Enc} & \textbf{Dec} & \textbf{Cpy} & \textbf{Instructions} \\ \midrule
1 & Producer-Consumer & $L$ & $L$ & $3 \cdot L$ & $2$ \\
2 & Proxy & $2 \cdot L$ & $2 \cdot L$ & $6 \cdot L$ & $4$ \\
3 & Client-Server & $L$ & $L$ & $3 \cdot L$ & $2$ \\ \bottomrule
\end{tabular}
}
\caption{Data sharing overheads of spatial isolation vs.
\codename. $L$: data size (memory words) in the shared region.
}
\label{tab:overheads_summary}
\end{table}
 
\paragraph{Summary.} The spatial \shmem baseline requires 
multiple data copies (see Figure~\ref{fig:problems}) to avoid attacks from the OS. Table~\ref{tab:overheads_summary} summarizes the encrypt/decrypt and copy operations incurred in our presented data patterns, assuming a region of $L$ memory words is shared and each word
is accessed once.

\subsection{Problem Formulation}
\label{sec:desired-properties}

The spatial isolation forces a rigid memory model. The type of
permissions of a memory region cannot change over time. The authority
which controls the permissions is fixed, i.e., the OS for public
memory and an enclave for private memory, regardless of the trust
model desired by the application.
We ask the following research question: {\em Does there exist a
minimal relaxation of the spatial model, which retains its
security guarantees, while eliminating its performance bottlenecks?}

\paragraph{Security Model.}
We assume that the OS is arbitrarily malicious and untrusted. The target application is partitioned into enclaves, which share one or more regions of memory. Each enclave has a {\em pre-agreed set of permissions}, which the application desires for its legitimate functionality. This set 
does not change, and in a sense, is the maximum permissions an enclave
needs for that region at any point of time in the region's lifetime. 
Any subset of enclaves can become compromised during the execution. We
refer to compromised enclaves as {\em faulty} which can behave arbitrarily. 
While providing better performance, there are $2$ security properties we desire from our TEE. First, the TEE interface does not allow faulty (and non-faulty) enclaves to escalate their privileges beyond the pre-agreed set. The second property, loosely speaking, ensures that faulty enclaves cannot obtain access permissions to the shared region, i.e., outside of the sequence that non-faulty enclaves desire to enforce. We detail these $2$ properties in  Section~\ref{sec:security_discussion}.
We additionally desire two soft goals.

\paragraph{Goal 1: Flexibility vs. Security.}
We aim to design a TEE memory model that offers security
comparable or better than our proposed baseline. A naive design,  which
allows unfettered flexibility to control a region's permissions, can
expose enclaves to a larger attack surface than the baseline. Enclaves
may maliciously compete to become an arbiter of permissions for a
region. It may be difficult to enforce a single consistent global view
of the permissions that each enclave has to a shared region, if
permissions can change dynamically. This in turn may create TOCTOU
bugs, since enclaves may end up making trust decisions based on a
stale view of another enclave's current permissions.
Therefore, our goal is to strike a balance between flexibility
and security.

\paragraph{Goal 2: Minimizing Implementation Complexity.} 
Enabling a new memory model may incur significant
implementation complexity. A complex memory model could introduce
expensive security metadata in hardware, increase the number of
signals, and introduce a large number of instructions. These can
directly lead to performance bottlenecks in the hardware
implementation, or have an unacceptable cost in chip area or power
consumption. Our goal is thus to keep the memory model {\em simple}
and minimize implementation complexity.

\paragraph{Scope.}
The TEE implementation is assumed to be bug-free. We aim to provide integrity,
confidentiality, and access control for shared memory data. We do not
aim to provide availability hence denial-of-service (DoS) attacks on
shared memory are not in-scope, since the OS may legitimately kill an
enclave or revoke access to memory. Further, our focus is on defining
a memory interface---micro-architectural implementation flaws and
side-channels are out of our scope. Lastly, if the TEE wishes to
safeguard the physical RAM or bus interfaces, it may require
additional defenses (e.g., memory encryption), which
are orthogonal and not considered here.
 \begin{figure*}[!tb]
\centering
    \subfloat[Producer-consumer (two-way
    isolated) \label{fig:prod-cons-soln2}]{
        \raisebox{-0.5\height}{\includegraphics[scale=0.325]{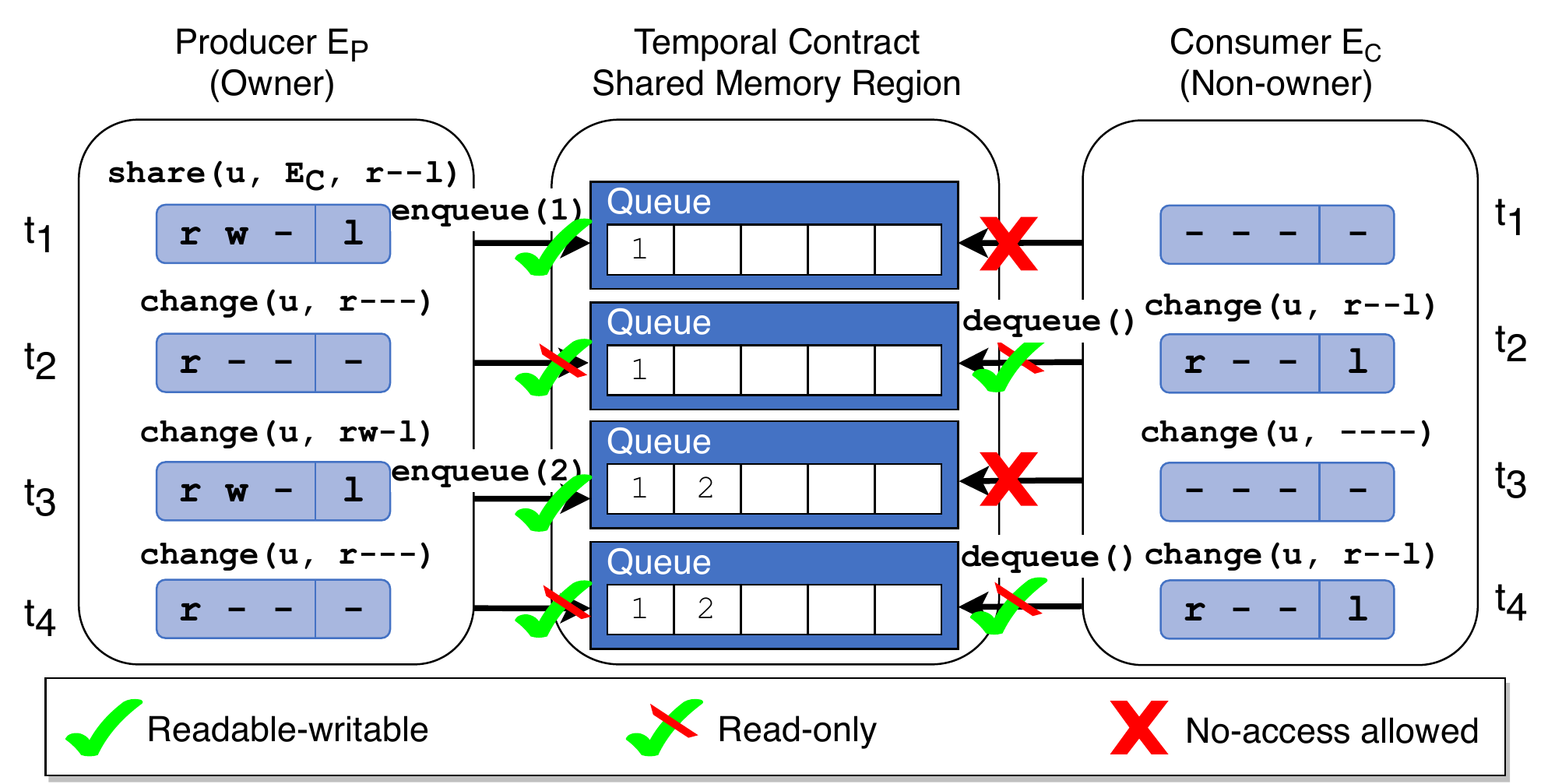}}
    }
    \hfill
    \subfloat[Proxy\label{fig:proxy-soln}]{
        \raisebox{-0.5\height}{\includegraphics[scale=0.325]{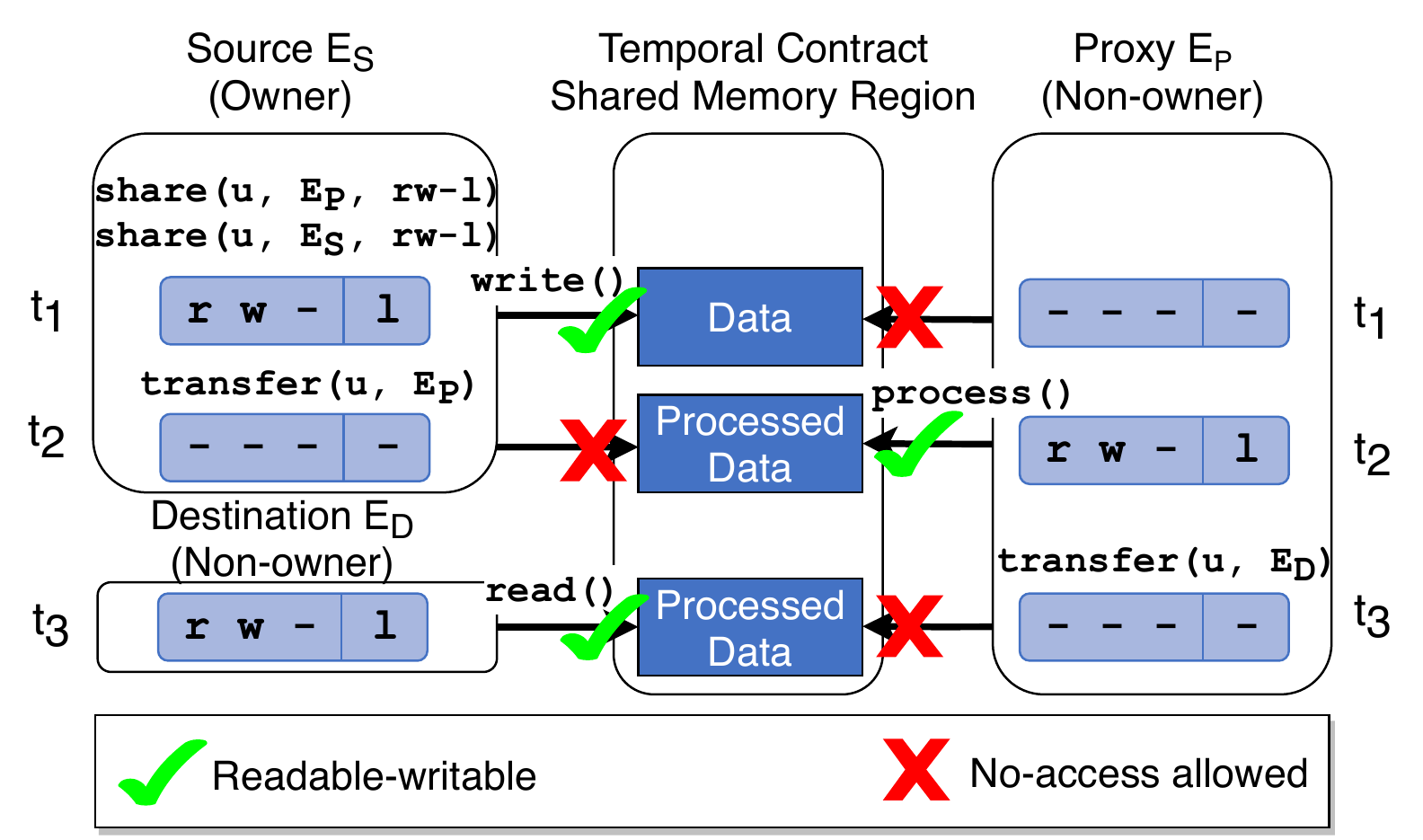}}
    }
    \hfill
    \subfloat[Client-server\label{fig:client-server-soln}]{
        \raisebox{-0.5\height}{\includegraphics[scale=0.325]{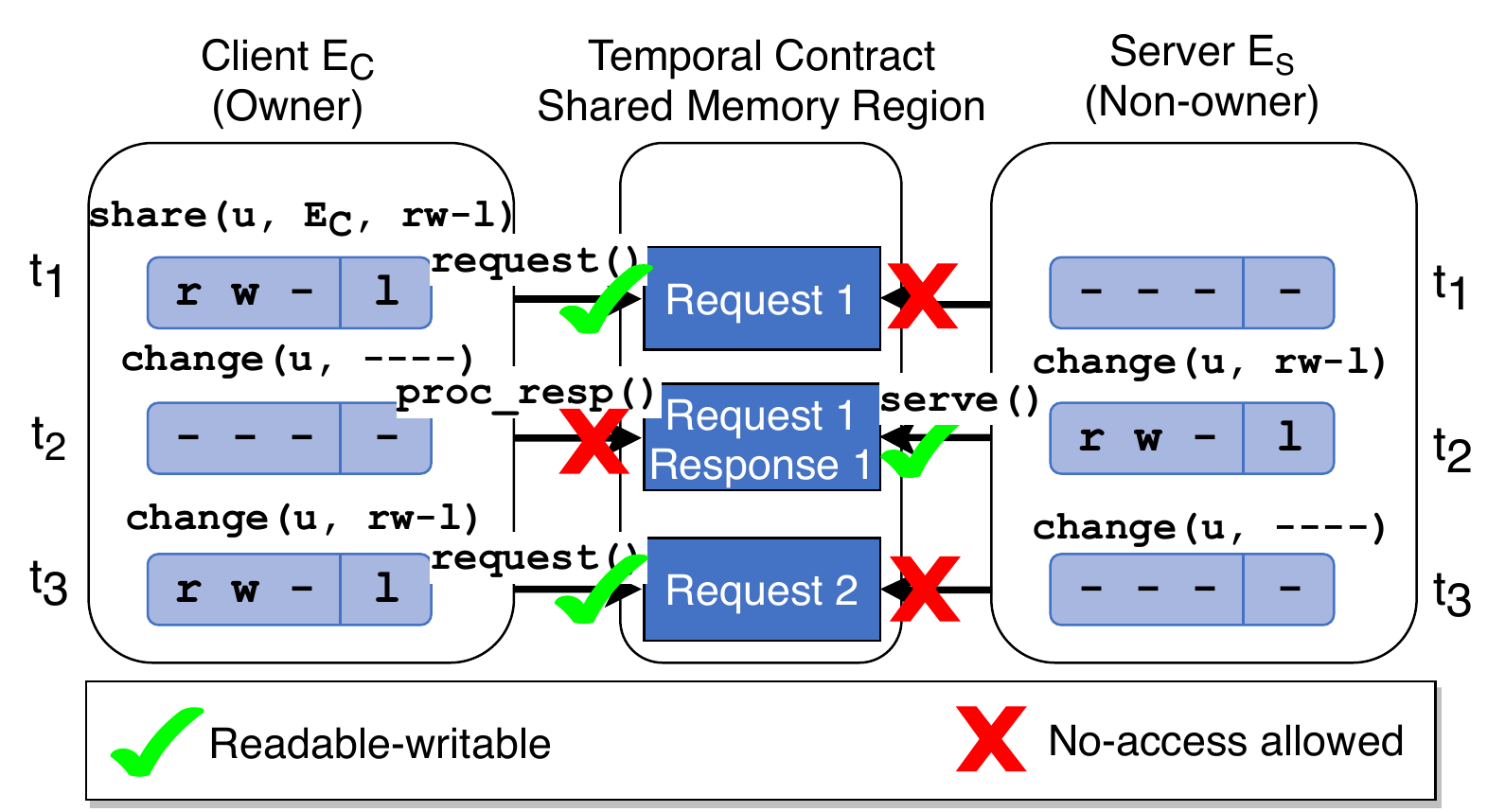}}
    }
\caption{Data sharing patterns with \codename.
}
\label{fig:data-patt-solutions}
\end{figure*}

\section{The \codename Memory Interface}
\label{sec:memory-interface}
\codename is a relaxation of the spatial isolation model i.e., 
It allows enclaves to share memory regions more
flexibly.

\subsection{Overview}
\label{sec:overview}

\codename
highlights the importance of $3$ key first-class abstractions that allow interacting enclaves to: 
(a)~have individual {\em asymmetric
permission views} of shared memory regions, i.e., every enclave can
have their local view of their memory permissions;
(b)~{\em dynamically} change these permissions as long as
they do not exceed a pre-established maximum; and
(c)~obtain {\em exclusive} access rights over shared memory regions,
and transfer it atomically in a controlled manner to other enclaves.

As a quick point of note, we show that the above three abstractions
are sufficient to drastically reduce the performance costs highlighted
in Section~\ref{sec:pattern-costs}. In particular,
Table~\ref{tab:overheads_summary} shows that in \codename, the number
of instructions is a small constant and 
 the number of data copies reduces to $1$ in all cases.
Whereas the spatial
\shmem baseline requires operations linear in the size $L$ of the
shared data accessed. We
will explain why such reduction is achieved in
Section~\ref{sec:patterns-solution}. But, in a brief glance at Figure~\ref{fig:data-patt-solutions} shows how the $3$ patterns can be implemented with a single shared memory copy, if the abstractions (a)-(c) are available. Notice how enclaves require different permission limits in
their views, which need to be exclusive sometimes, and how permissions change over time. For security, it is necessary that accesses made by enclaves are serialized in particular (shown) order.

Our recommendation for the $3$ specific \codename abstractions is
intentional. Our guiding principle is simplicity and security---one could easily relax the spatial memory model further, but this comes at the peril of subtle security issues and increased implementation complexity. We discuss these considerations in Section~\ref{sec:security_discussion} after our design details. 

\begin{figure*}[!ht]
\centering
    \raisebox{-0.5\height}{\begin{minipage}[b]{0.68\linewidth}
{ \footnotesize
\begin{tabularx}{\linewidth}{XXX}
\toprule
\textbf{Instruction} &
    \textbf{Permitted Caller} & \textbf{Semantics}\\
\midrule
        \texttt{uid =
        create(size)} & owner of \texttt{uid} & create a  region \\
\texttt{err = map(vaddr, uid)} & accessor of \texttt{uid} & map
    VA range to a  region \\
\texttt{err = unmap(vaddr, uid)} & accessor of \texttt{uid}  & remove
     region mapping \\
\texttt{err = share(uid, eid, P)} & owner of \texttt{uid} &
    share region with an
        enclave \\
\texttt{err = change(uid, P)} & accessor of
    \texttt{uid} & adjust the actual access permission to a memory
    region \\
\texttt{err = destroy(uid)} & owner of \texttt{uid} & destroy a memory
    region\\ 
\texttt{err = transfer(uid, eid)} & current lock holder  & transfer lock to another
    accessor \\
\bottomrule
\end{tabularx}
    }
\captionof{table}{Summary of security instructions in \codename.}
\label{tab:instructions}
    \end{minipage}}
    \hfill
    \raisebox{-0.5\height}
    {\begin{minipage}[b]{0.3\linewidth}
\centering
\includegraphics[width=0.65\linewidth]{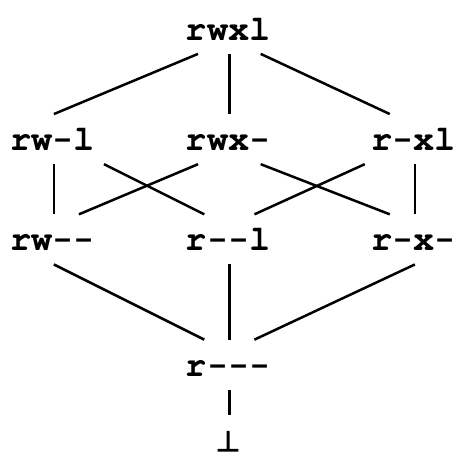}
        \captionof{figure}{Lattice for the permission hierarchy, or $\leq$ relation for permissions.}
        \label{fig:permission_lattice}
    \end{minipage}}
\end{figure*}

\subsection{\codename Abstractions}

\codename relaxes spatial isolation by allowing enclaves to define and
change permissions of  regions shared externally. Our design works at
the granularity of {\codename} {\em memory regions}. These 
regions are an abstraction defined by our model; each region maps to a
contiguous range of virtual memory addresses in the enclave. From the view of each enclaves, an \codename memory region has $4$ permission bits: standard {\em read}, {\em
write}, {\em execute}, and a protection {\em lock} bit.

For each memory region, we have two types of enclaves.
The first are {\em owners}, who have the sole privilege to create,
destroy, and initiate sharing of regions. The second kind of enclaves
are {\em accessors}. Owners can share and grant the permission to
accessors only for the regions they own. An enclave can be both an
owner and an accessor of a region. 

\codename gives $3$ first-class abstractions: {\em
asymmetry}, {\em dynamicity}, and {\em exclusivity} in an enclave's permission
views.

\paragraph{Asymmetric Permission Views.} 
In several data patterns shown in Section~\ref{sec:pattern-costs}, 
one can see that different enclaves require different permissions of
the shared memory. For example, one enclave has read accesses whereas
others have write-only access. The spatial model is a ``one size fits
all'' approach that does not allow enclaves to setup asymmetric
permissions for a public memory region securely---the OS can always
undo any such enforcement that enclaves might specify via normal
permission bits in hardware.
In \codename, different enclaves are allowed to specify their own set
of permissions (or {\em views}) over the same shared region, which are
enforced by the TEE. This directly translates to avoiding the costs of
creating data copies into separate regions, where each region has a
different permission.
For example, in Pattern 1 the producer has read-write permissions and
the consumer has read-only permissions for the shared queue.

\paragraph{Dynamic Permissions.}
In \codename, enclaves can change their permissions over time, 
without seeking consent from or notifying other enclaves. In the
spatial isolation model, if enclaves need different permissions over
time on the same shared data, separate data copies are needed.
\codename eliminates the need for such copies.
For example, in Pattern 2, when the source enclave generates data it
has read-write permissions, while the proxy enclave has no
permissions. After that, the source drops all its permissions, and
proxy enclave gains read-write permissions to process the data. This
way, both source and proxy enclaves do not interfere with each others
operations on the shared region.

While enabling dynamic permissions, \codename does not allow enclaves
to arbitrarily escalate their permissions over time. In \codename,
only the owner can share a memory region with other accessors during the
lifetime of the memory region. When the owner shares a memory region, it sets
the {\em static maximum} permissions it wishes to allow for the
specified accessor at any time. This static maximum  cannot be changed
once set by the owner for a specified enclave. Accessors can escalate
or reduce their privileges dynamically. But if the accessor tries to
exceed the static maximum at any given point in time, \codename
delivers a general protection fault to the accessor enclave.

\paragraph{Exclusive Locks.}
\codename incorporates a special bit for each memory region called the
{\em lock} bit. This bit serves as a synchronization mechanism between
enclaves, which may not trust each other. \codename ensures that at
any instance of time only one accessor has this bit set, thereby
enforcing that it has an exclusive access to a region. When this bit
is on, only that accessor is able to access it---all other accessors
have their permissions temporarily disabled. When the lock is acquired
and released by one enclave, all accessors and the owner of that
region are informed through a hardware exception/signal. Lock holders
can release it generically without specifying the next holder or can
atomically transfer the lock to other accessors through  \transfer
instruction. Atomic transfers become useful for flexible but
controlled transfer of exclusive access over regions. For example, in
Pattern 2, the source holds the lock bit for exclusive access to the
region for writing its request. Thus, no one can tamper with the
packet while the source writes to it. 
Then, the source transfers the lock directly
to the proxy. Proxy exclusively accesses the region to update the
packet and then transfers the lock to the destination. Only then the  destination can read 
the updated packet.

\subsection{Design Details}
\label{sec:design}

\codename is a memory interface specification consisting of $7$
instructions, summarized in Table~\ref{tab:instructions}, which
operate on \codename memory regions. Each \codename region is
addressable with a {\em universal} identifier that uniquely identifies
it in the global namespace. Universal identifiers can be mapped to
different virtual addresses in different enclaves, and at the same
time, are mapped to physical addresses by a \codename implementation.
The \codename interface semantics are formalized as pre- and post-conditions
in Appendix~\ref{sec:appx}, which any secure implementation of this interface should satisfy. 
Next, we explain the
\codename design by walking through the typical lifecycle of a region.

\paragraph{Owner View.} 
Each memory region $r$ has a unique owner enclave throughout its
lifetime. 
An enclave $p$ can create a new memory region $r$ with  \create instruction, which takes the memory region size
and returns a universal id (\uid). The enclave $p$ is the owner of
the new memory region $r$. 
The owner permissions are initialized to an owner-specified safe
maximum. These permissions
are bound to a
memory region. The owner, just like any accessor, can bind the memory
region to its virtual address space by using \map and \unmap
instructions. The \map instruction takes a \uid for the region
and the virtual address range to map to it. A memory region can be
mapped at multiple virtual address in an enclave, but the static
permissions bound to the region at the time of creation apply to all
instances mapped in the virtual address space.
The owner can then share the memory with any other enclave using the
\share instruction, which specifies the \uid of the memory
region, the enclave id of the other accessor, and the static maximum
permissions allowed for that accessor. 

Every accessor, including the owner, can dynamically change the
permissions of a memory as long as the permissions are strictly more
restrictive (fewer privileges) than the static maximum for the
enclave. For the owner, the static maximum is the full set of
permissions, and for other accessors, it is determined by the \share
instruction granting the access. The changes to such permissions are
local to each accessor, i.e., permission changes are not globally
effected for all accessors; rather they apply to each enclave
independently. The lattice shown in
Figure~\ref{fig:permission_lattice} defines the permission hierarchy.
Finally, the owner can destroy the memory region at any point in time
by invoking the \destroy instruction. \codename sends all accessors a
signal when the owner destroys a memory region. Destroying a region
ends the lifetime in all enclaves. The OS can invoke the \destroy
instruction on an enclave to reclaim the memory region or to protect
itself from denial-of-service via the enclave.

\paragraph{Accessor's View.}
The accessor binds a memory region in its virtual address space using
the \map instruction; the same way as owners do. The initial
permissions of the memory region are set to static maximum allowed by
the owner (specified by the owner in the \share instruction). The
accessor can restrict its permissions dynamically further at any time
as long as the resulting permissions are below this static maximum
using the \change instruction. Such changes, as mentioned previously,
remain locally visible to the accessor enclave.

\paragraph{Permission Checks.}
The \codename TEE implementation enforces the permissions defined by
enclaves in their local views. A permission bit is set to $1$ if the
corresponding memory access (read, write, or execute) is allowed, and
set to $0$ otherwise. For memory accesses, the security checks 
can be summarized by two categories:
(1) availability check of the requested resources (e.g., memory
regions and enclaves), which ensures that instructions will not be
performed on non-existing resources; and (2) permission checks of the
caller, which ensures that the caller has enough privilege to perform
the requested instruction. Table~\ref{tab:instructions} defines the permitted caller for each instruction. For example, \share and \destroy
instructions can only be performed by the owner of the region. 

The \change instruction is the mechanism for dynamically updating
permissions of a \codename region. \codename requires that the newly requested permissions ($perm$) by an enclave fall within the limits of its static maximum permissions ($max$).
Specifically, \codename checks that $perm \leq max$, where
the $\leq$ relation is defined by the lattice shown in Figure~\ref{fig:permission_lattice}. The lock bit
can only be held (set to $1$) in the local view of a single
enclave at any instance of time. When it is set for one enclave,
that enclave's local permission bits are enforced, and
all other enclaves have no access to the region.
When lock is set to $0$ in the local views of all enclaves, permissions of each enclave are as specified in its local view.

\begin{figure}[!tb]
\end{figure}

\paragraph{Lock Acquire \& Release.} 
Accessors can attempt to ``acquire'' or ``release'' the lock by using
the \change instruction. It returns the accessor's modified
permissions, including the lock bit that indicates whether the acquire
/ release was successful. \codename ensures that at any instance of
time, only a single enclave is holding the lock. If any other enclave
accesses the region or tries to issue a \change instruction on that
region's permissions, these requests will be denied. 

A lock holder can use the \change instruction to release locks;
however, there are situations where the holder wishes to explicitly
specify who it intends to be next holder of the lock. \codename allows
lock holder to invoke a \transfer instruction which specifies the
enclave id of the next desired accessor. The next holder must have the
memory region mapped in its address space for the transfer to be
successful. 

\paragraph{\codename Exceptions \& Signals.}
\codename issues exceptions whenever memory operations violating any
permission checks are made by an enclave. \codename notifies enclaves
about events that affect the shared memory region via asynchronous
signals.
Signals are issued under two scenarios.
First, when the owner destroys a memory region $r$, permissions
granted to other enclaves will be invalidated since the memory region
is not in existence. In order to prevent them from continuing without
knowing that the memory region can no longer be accessed, the security
enforcement layer will send signals to notify all accessors who had an
active mapping (i.e., mapped but not yet unmapped) for the destroyed
memory region.
The second scenario for signals is to notify changes on lock bits. 
Each time an accessor successfully acquires or releases the lock
(i.e., using \change or \transfer instructions), a signal is issued to
the owner. The owner can mask such signals if it wishes to, or it can
actively monitor the lock transfers if it desires. When a transfer
succeeds, the new accessor is notified via a signal.

Lastly, we point out that \codename explicitly does not introduce
additional interface elements, for example, to allow enclaves to
signal to each other about their intent to acquire locks, or to prevent starvation. Section~\ref{sec:security_discussion} discusses these considerations to avoid interface complexity.

\subsection{Performance Benefits}
\label{sec:patterns-solution}

\codename relaxes the spatial isolation model by introducing
flexibility in specifying permissions over shared regions. We now
revisit the example patterns discussed in
Section~\ref{sec:pattern-costs} to show these patterns can be
implemented with significantly lower costs (summarized in
Table~\ref{tab:overheads_summary}) with \codename.

\paragraph{Revisiting Pattern 1: Producer-Consumer.}
Application writers can partition the producer and consumer into two
enclaves that share data efficiently with \codename. We can consider
two scenarios of faulty enclaves. The first allows one-way protection,
where the producer safeguards itself from a faulty consumer. The
second offers mutual protection where faults in either enclave do not
spill over to the other.

\begin{figure}[!tb]
\centering
\includegraphics[width=0.8\linewidth]{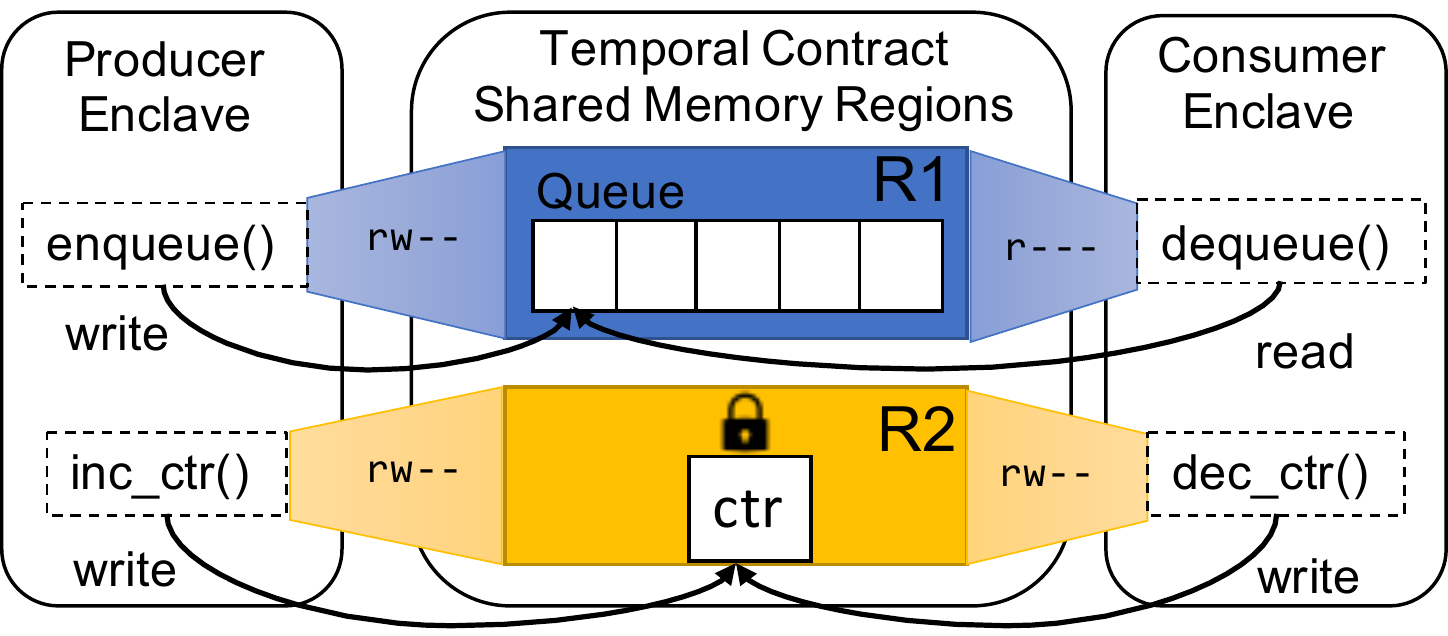}
\caption{One-way isolated producer-consumer pattern with \codename. 
The producer writes to a memory region (R1) shared with consumer; consumer
is only allowed to read.
}
\label{fig:prod-cons-soln1}
\end{figure}

In the one-way isolation scenario, the producer can create a memory
region and share it with the consumer with the maximum permission set
to $\tt{r{}-{}-{}-}$. The producer and the consumer can then keep
their permissions to $\tt{r{}w{}-{}-}$ and $\tt{r{}-{}-{}-}$
respectively, which allow the producer to read and write data and the
consumer to only read data in the memory region 
(Figure~\ref{fig:prod-cons-soln1}). The producer can directly write
its data to the shared memory region, and the consumer can directly
read from it without needing to moving the data back and forth between
the shared memory region and their private memory. The producer can
ensure that the consumer, even if compromised, cannot maliciously race
to modify the data while it is being updated in a critical section by
the producer. The whole process does not involve any extra data copies
or a cryptographically secure public memory, and only introduces fixed
costs of setting up and destroying the memory regions.

Two-way isolation is desired when both producer and consumer wish to
modify shared data in-place, while assuming that the other is faulty.
As a simple example, counters in shared queue data structures often
require atomic updates. In general, the producer and consumer may want
to securely multiplex their access to any amount of shared data (e.g.,
via shared files or pipes) for performing in-place updates. \codename
makes this possible without creating any additional memory copies or
establishing secure channels. The shared memory region can be created
by (say) the producer and shared with the consumer with a static
maximum permission of $\mathtt{r{}w{}-{}l}$ as shown in
Figure~\ref{fig:prod-cons-soln2}. When either of them wish to acquire
exclusive access temporarily, they can use the \change instruction,
setting it from $0$ to $1$. Therefore, the only overhead incurred is
that of the execution of the \change instruction itself, which is in
sharp contrast to the $2$ copies of the entire shared data required in
spatial isolation model.

\paragraph{Revisiting Pattern 2: Proxy.} 
The proxy example can be seen as a straight-forward sequential
stitching of two producer-consumer instances. The shared data would
first be written by the producer, then the proxy atomically reads or
updates it, and then the consumer would read it. All three entities
can hold the lock bit in this order to avoid any faulty enclave to
access the shared memory where unintended. 
\codename transfer instruction eliminates windows of attack when
passing the locks from one enclave to another. 
Specifically, it allows the source to atomically transfer the lock to
proxy, who then atomically transfers it to the consumer. In this way,
the proxy workflow can be implemented without any extra copy of the
shared data as shown in Figure~\ref{fig:proxy-soln}.

\paragraph{Revisiting Pattern 3: Client-Server.} 
The client-server workflow can similarly be executed by keeping a
single copy of the shared data, as shown in
Figure~\ref{fig:client-server-soln}, which reduces the number of data
copies from $6$ in the case of spatial isolation to $1$.

\paragraph{Compatibility with Spatial Isolation.} 
It is easy to see that \codename is strictly more expressive than
spatial isolation model, and hence keeps complete compatibility with
designs that work in the spatial isolation model. Setting up the
equivalent of the public memory is simple---the owner can create the
region and share it with $\mathtt{r{}w{}x{}-}$ for all. Private memory
simply is not shared after creation by the owner.

\paragraph{Privilege De-escalation Built-in.}
In \codename, enclaves can self-reduce their privileges below the
allowed maximum, without raising any signals to other enclaves. This
design choice enables compatibility with several other low-level
defenses which enclaves may wish to deploy for their own internal
safety---for example, making shared object be non-executable, or
write-protecting shared security metadata.

 \subsection{Security \& Simplicity}
\label{sec:security_discussion}

We begin by observing that it is straight-forward to implement the
\codename interface with an (inefficient) trusted enforcement layer
using spatially isolated memory\footnote{The
enforcement layer would be implemented by a trusted enclave, which
would keeps the shared memory content and the permission matrix in its
{\em private} memory. Each invocation of a \codename instruction 
would translate to a RPC call to the enforcement enclave, which could
simply emulate the stated checks in Table~\ref{tab:instructions} and
Appendix~\ref{sec:appx} as
checks on its matrix.}. It follows that any memory permission
configurations which may be deemed as ``attacks'' on the \codename
would also be admissible in a faithful emulation on the spatial
isolation model. In this sense, \codename does {\em not degrade
security} over the spatial isolation model. The primary highlight of
\codename is the performance gains it enables without degrading security.

We point out two desirable high-level security properties that
immediately follow from the \codename interface
(Table~\ref{tab:instructions}). Application writers can rely on these
properties without employing any extra security mechanisms.

\paragraph{Property 1: Bounded Escalation.}
If an owner does not explicitly authorize an enclave $p$ access to a
region $r$ with a said permission, $p$ will not be able to obtain that
access.

This property follows from three design points:
(a)~Only the owner can change the set of enclaves that can access a
region. Non-owner enclaves cannot grant access permissions to other
enclaves since there is no permission re-delegation instruction in the
interface.
(b)~Each valid enclave that can access a region has its permissions
bounded by an owner-specified static maximum. 
(c)~For each access or instruction, the accessor enclave and the
permission is checked to be legitimate by each instruction in the
interface (see Table~\ref{tab:instructions}).

\paragraph{Property 2: Enforceable Serialization of Non-faulty Enclaves.}
If the application has a pre-determined sequence in which accesses of
non-faulty enclaves should be serialized, then \codename can guarantee
that accesses obey that sequence or will be aborted. Specifically, let
us consider any desired sequence of memory accesses on an \codename
region $a_1,a_2,\dots,a_n$ and assume that all enclaves performing
these accesses are uncompromised.
Then, using \codename, the application writer can guarantee that its
sequence of accesses will follow the desired sequence, even in the
presence of other faulty enclaves, or can be aborted safely.

The property can be enforced by composing two \codename abstractions:
(a)~For each access $a_i$ by an enclave $e(a_i)$ in the pre-determined
sequence, the accessor can first acquire the lock to be sure that no
other accesses interfere.
(b)~When the accessor changes, say at access $a_j$, the
current enclave $e(a_j)$ can safely hand-over the lock to the next
accessor $e(a_{j+1})$ by using the \transfer instruction. Faulty
enclaves cannot acquire the lock and access the region at any
intermediate point in this access chain. For example, in 
Pattern 2 (proxy), once the proxy enclave modifies
the data in-place, simply releasing the lock is not {\em safe}. A
faulty source enclave can acquire the lock before the destination does
and tamper with the data. However, the proxy can eliminate such an
attack from a racing adversary using the \transfer instruction.

\paragraph{Simplicity.}
Several additional design decisions make \codename simple to reason
about. We discuss two of these: forcing applications to use {\em
local-only views} in making trust decisions and {\em minimizing
interface complexity}.

Each enclave is asked to make security decisions based only on its own
{\em local} view of its current and static maximum permissions. This
is an intentional design choice in \codename to maintain simplicity. One
could expose the state of the complete access control (permission)
matrix of all enclaves to each other, for enabling more flexible or
dynamic trust decisions between enclaves. 
However, this would also add complexity to application writers.
All enclaves would need to be aware of any changes to the global
access permissions, and be careful to avoid any potential TOCTOU
attacks when making any trust decisions based on it.
To simplify making trust decisions, the only interface in \codename
where an enclave assumes a global restriction on the shared memory is
the lock bit. When using this interface, the lock holder can assume
that it is the only holder of the lock globally. 

\codename admits a simpler TEE implementation. The interface avoids
keeping any metadata that changes based on shared memory state (or
content). The TEE hardware can keep all security-relevant metadata in
access control tables, for instance. Since enclaves do not have
visibility into the global view of permissions of all other enclaves,
the TEE does not need to set up additional mechanisms to notify
enclaves on changes to this table (e.g., via signals). Further,
\codename does {\em not} provide complete transaction memory
semantics, i.e., it does not provide atomic commits or memory
rollbacks, which come with their own complexity~\cite{rote}.

Similarly, consider starvation: A malicious or buggy
enclave may not release the lock. A more complex interface than
\codename would either require the TEE to arbitrate directly or allow
the owner to do so, say to have memory be released within a time
bound. 
However, such a solution would come with considerable interface
complexity. It would open up subtle attack avenues on the lock holder.
For instance, the enclave could lose control of the shared memory when
its contents are not yet safe to be exposed. 
Instead, \codename simply allows owners to be notified when enclaves
issue requests to acquire locks via the \change instruction. Enclaves
can implement any reasonable policy for starvation---for example, to
tear down the memory securely if a lock holder is unresponsive or
repeatedly acquiring locks to the region.
  \section{Implementation on \riscv}
\label{sec:implementation}

We build a prototype implementation of \codename on an open-source
 RocketChip quad-core SoC~\cite{riscvpriv, rocket}. We utilize $2$
building blocks from the \riscv architecture, namely its physical
memory protection (PMP) feature and the programmable machine-mode
($\tt{m-mode}$). Note that \codename does not make any 
changes to the hardware. 
We use Keystone---an  open-source framework for
instantiating new security TEEs such as
\codename~\cite{dayeol2020keystone}. Keystone provides a Linux driver
and a convenient SDK to create, start, resume, and terminate enclaves
by using the aforementioned features. It supports  
$\tt{gcc}$ compiler and has C/C++ library to 
build enclave applications. Keystone originally uses the spatial
isolation model, which we do not.

\paragraph{\riscv PMP and $\tt{m-mode}$.}
The physical memory protection (PMP) feature of \riscv allows software
in machine-mode (the highest privilege level in \riscv) to restrict
physical memory accesses of software at lower privilege levels
(supervisor- and user-modes). Machine-mode software achieves this by
configuring PMP entries, which are a set of registers in each CPU
core. Each PMP register holds an entry specifying one contiguous 
physical address range and its corresponding access permissions.
Interested readers can refer to the \riscv standard PMP 
specifications~\cite{riscvpriv}.

The \codename TEE implementation runs as $\tt{m-mode}$ software. All the
meta-data about the memory regions, owners, static maximums, and the
current view of the permission matrix are stored in here. 
The $\tt{m-mode}$ is the only privilege-level that can modify PMP entries.
Thus, the OS ($\tt{s-mode}$) and the enclave ($\tt{u-mode}$) cannot read
or update any PMP entries or meta-data.
When the enclave invokes any \codename instruction, the execution traps 
and the hardware redirects it to the $\tt{m-mode}$. This control-flow
cannot be changed by $\tt{s-mode}$ or $\tt{u-mode}$ software.
After entering $\tt{m-mode}$, \codename first checks whether 
the caller of the instruction is permitted to make the call. If it is a valid entity who is permitted to invoke this instruction, \codename performs the meta-data, and if necessary, PMP updates.

\codename keeps two mappings in its implementation:
(a)~virtual address ranges of each enclave and the corresponding
\codename region universal identifier (uid); and (b)~the effective physical address range to which each uid is mapped.
Thus, when an enclave tries to access a virtual address, \codename
performs a two-level translation: from virtual address to a uid and
subsequently to the physical address.
The \map and \unmap instruction only require updating the
first mapping, as they update virtual to uid mappings only. The
\change, \share, and \transfer only update the second
mapping because they only change permission bits without affecting
virtual memory to uid bindings. The \create and \destroy 
instructions update both mappings. For enforcing access checks, the
\codename TEE additionally maintains a permissions matrix of the
current and the static maximum permissions of each enclave.
Permissions are associated with uids, not with virtual addresses.
For enforcement, the TEE translates uid permissions in the permission
matrix to physical memory access limits via PMP entries. Whenever the
permission matrix is updated by an enclave, the permission updates
must be reflected into the access limits in PMP entries. Further, one
PMP entry is reserved by \codename to protects is internal mappings
and security data structures.

The \riscv specification limits the number of PMP registers to $16$.
Since each region is protected by one PMP entry, this fixes the
maximum number of regions allowable across all enclaves simultaneously.
This limit is not due to \codename design, and one can increase the
number of PMP entries and \codename only increases the needed PMP
entries by $1$.

When context-switching from one enclave to another, apart from
the standard register-save restore, Keystone modifies PMP entries to
disallow access to enclave private memory---this is because it uses a
spatial isolation design. 
We modify this behavior to allow continued access to shared regions
even when the owner is not executing for \codename.
 \section{Evaluation} 
\label{sec:eval}

We aim at evaluate the following research questions:
\begin{itemize}
    \item How does the performance of \codename compare with  
    spatial \shmem baseline on  
    \riscv? 
    \item What is the impact of \codename on privileged software 
    trusted code base (TCB) and hardware complexity? 
\end{itemize}

We implement the spatial \shmem baseline and \codename on the same hardware
core, in order to singularly measure the difference between the
spatial and our \codename design.
Production-grade TEEs, such as Intel SGX, often have additional
mechanisms (e.g., hardware-level memory encryption, limited size of private physical memory, 
etc.) which are orthogonal to the performance gains due to our
proposed memory model. Our implementation and evaluation exclude these
 overheads.

\paragraph{Benchmarks.}
We experiment with $2$ types of benchmarks, using both our \codename
implementation and the described spatial \shmem baseline (Section~\ref{sec:baseline-model}) on the same \riscv core:
(a) simple synthetic programs we constructed that
implement the $3$ data patterns with varying number of regions and
size of data. We also construct synthetic thread synchronization
workloads with controllable contention for locks. 
(b) 
standard pre-existing real-world benchmarks, which include I/O 
intensive workloads (IOZone~\cite{iozone}), parallel computation 
(SPLASH-2~\cite{splash,splash-summary}), and CPU-intensive benchmarks  (machine 
learning inference with Torch~\cite{privado,
torch}). We manually modify these benchmarks to add \codename
instructions, since we do not presently have a compiler for \codename.

\paragraph{Experimental Setup.}
We use a cycle-accurate, FPGA-accelerated
(FireSim~\cite{DBLP:conf/isca/KarandikarMKBAL18}) simulation of
RocketChip~\cite{rocket}. Each system consists of $4$ RV64GC
cores, a 16KB instruction and data caches, $16$ PMP
entries per core (unless stated otherwise), and a
shared 4MB L2 cache. Area numbers were computed using a commercial
22nm process with Synopsys Design Compiler version L-2016.03-SP5-2
targeting \targetfreq{}. Other than varying the number of PMP entries, we
do not make any changes to RocketChip.

\subsection{Performance of \codename}
\label{sec:temporal_vs_spatial}

To evaluate the performance of \codename vs. spatial isolation, we
first used synthetic benchmarks that cover common types of data
sharing behaviors in applications, including the data sharing patterns
introduced in Section~\ref{sec:baseline-model}.

\begin{figure*}[!tb] \centering
\subfloat[Producer-consumer]{
\includegraphics[width=0.27\linewidth]{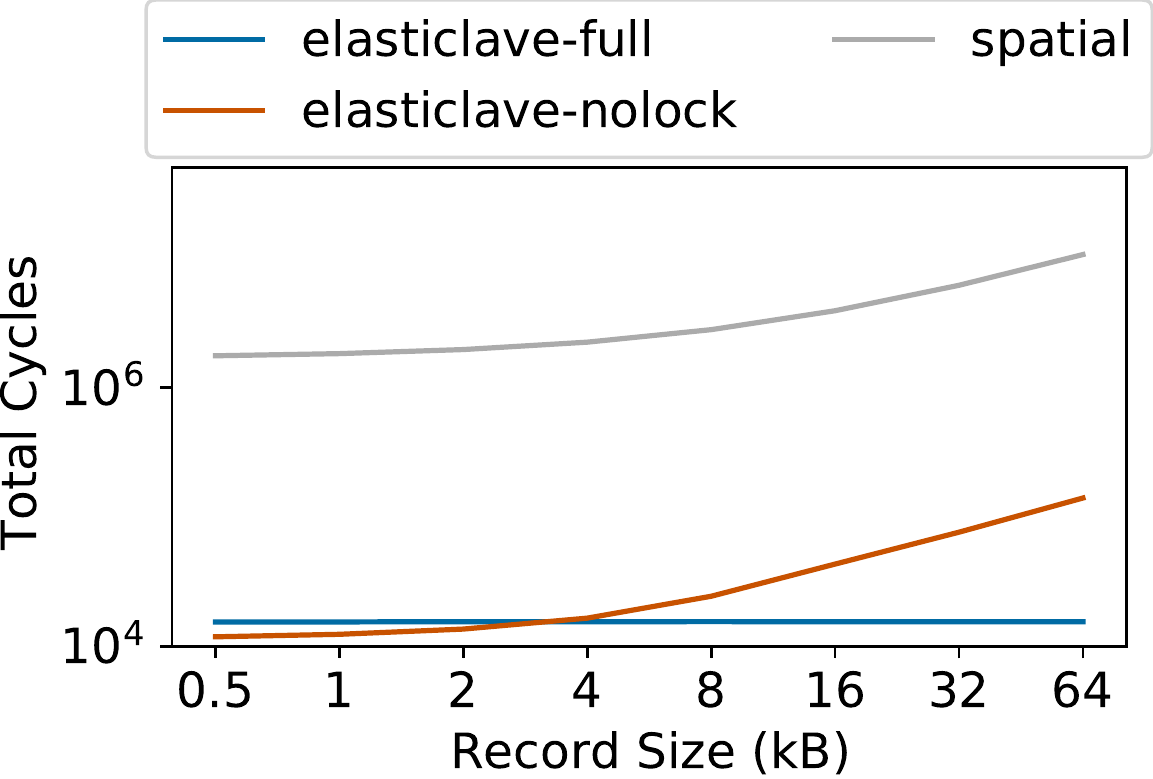}
}
\subfloat[Proxy]{
\includegraphics[width=0.27\linewidth]{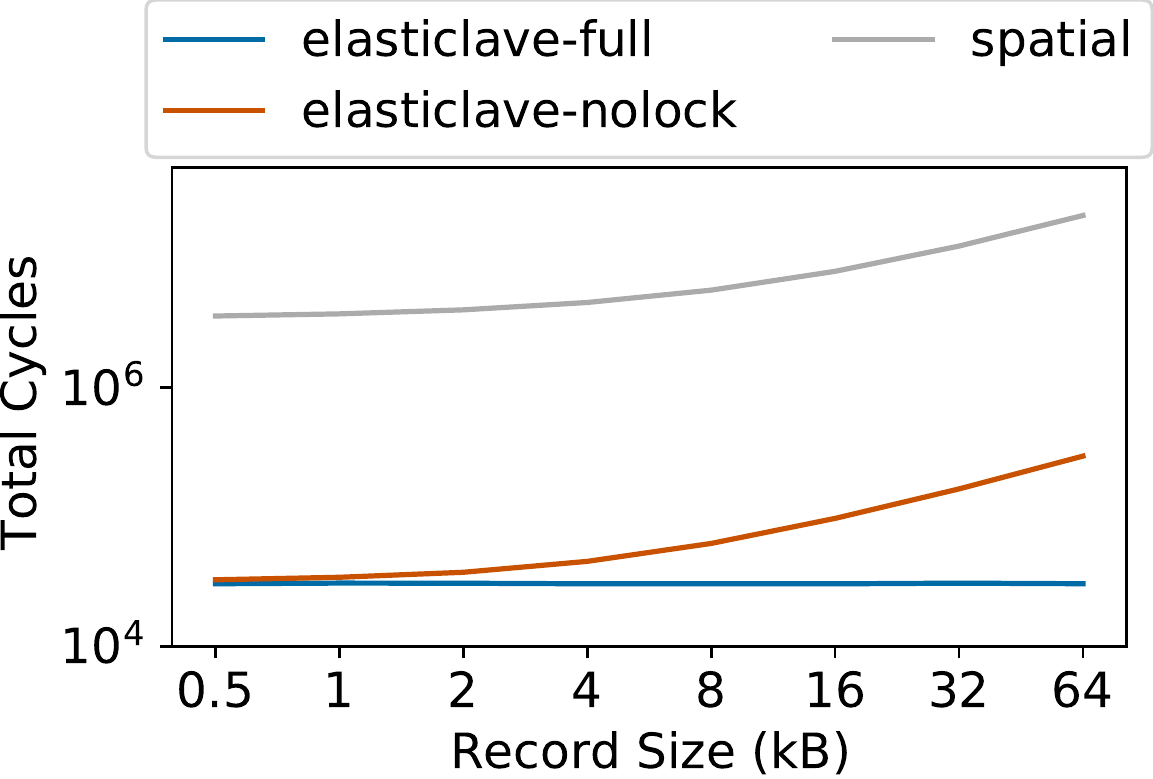}
}
\subfloat[Client-server]{
\includegraphics[width=0.27\linewidth]{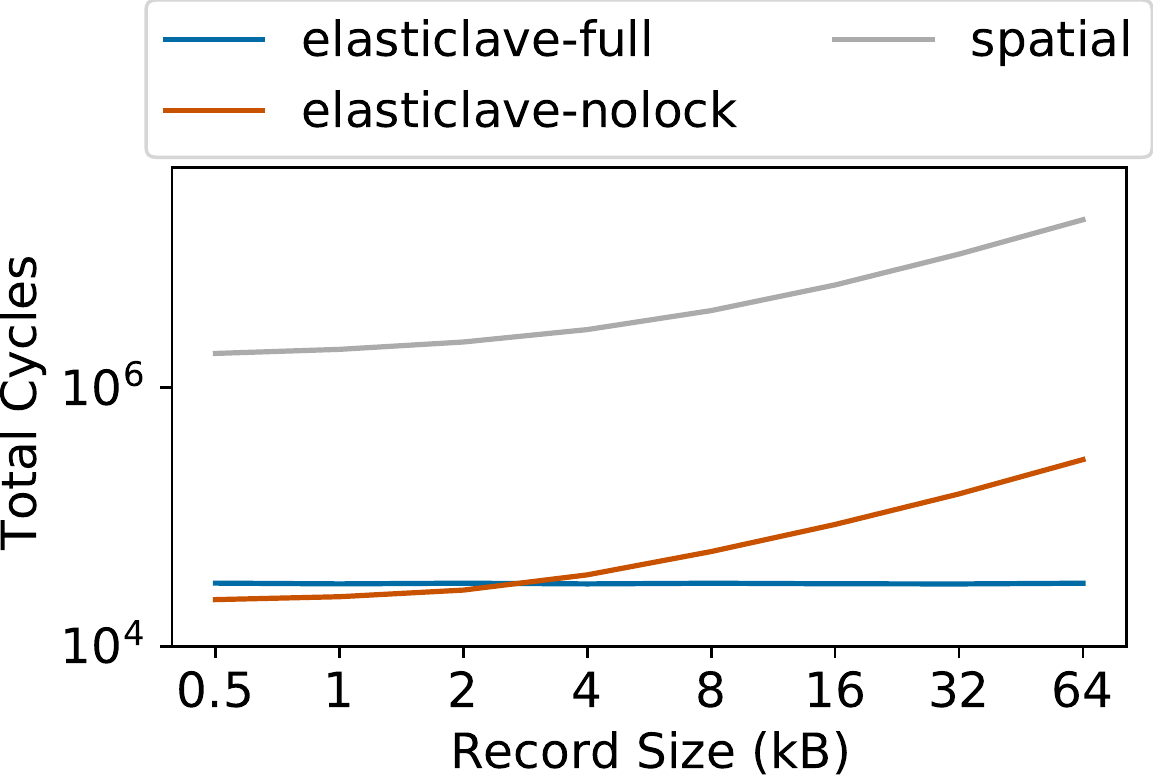}
}
\caption{Performance of the $3$ data-sharing patterns.}
\label{fig:icall_results} \end{figure*}

\begin{figure*}[!tb] \centering
\subfloat[\codename-full]{
\includegraphics[width=0.27\linewidth]{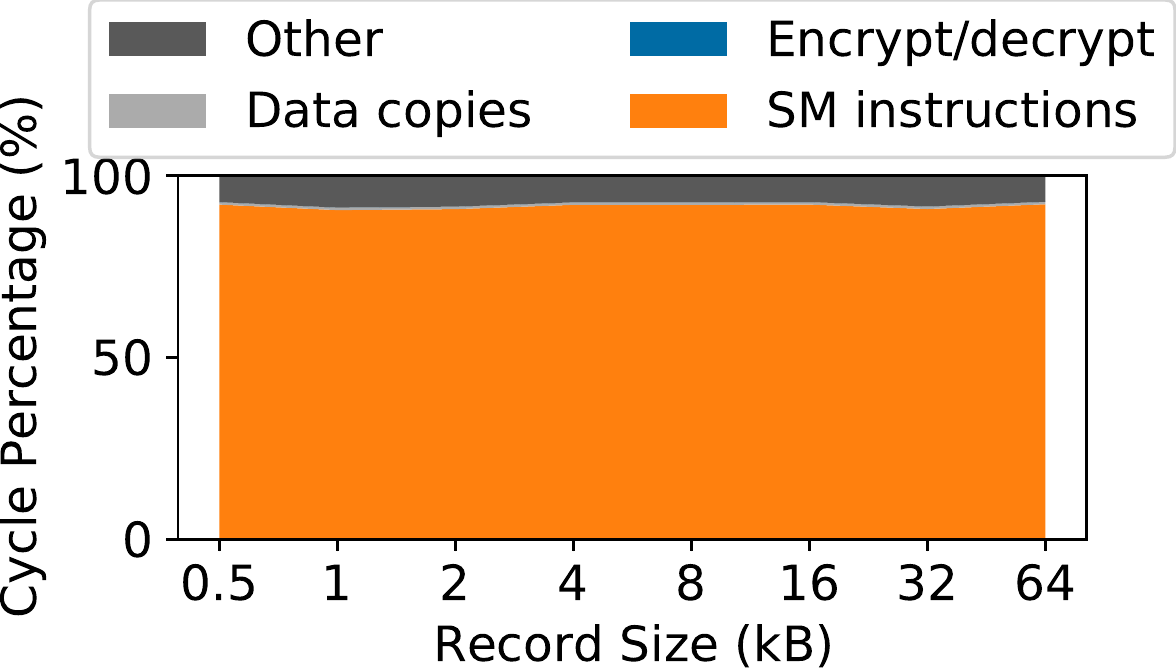}
}
\subfloat[\codename-nolock\label{fig:breakdown_nolock}]{
\includegraphics[width=0.27\linewidth]{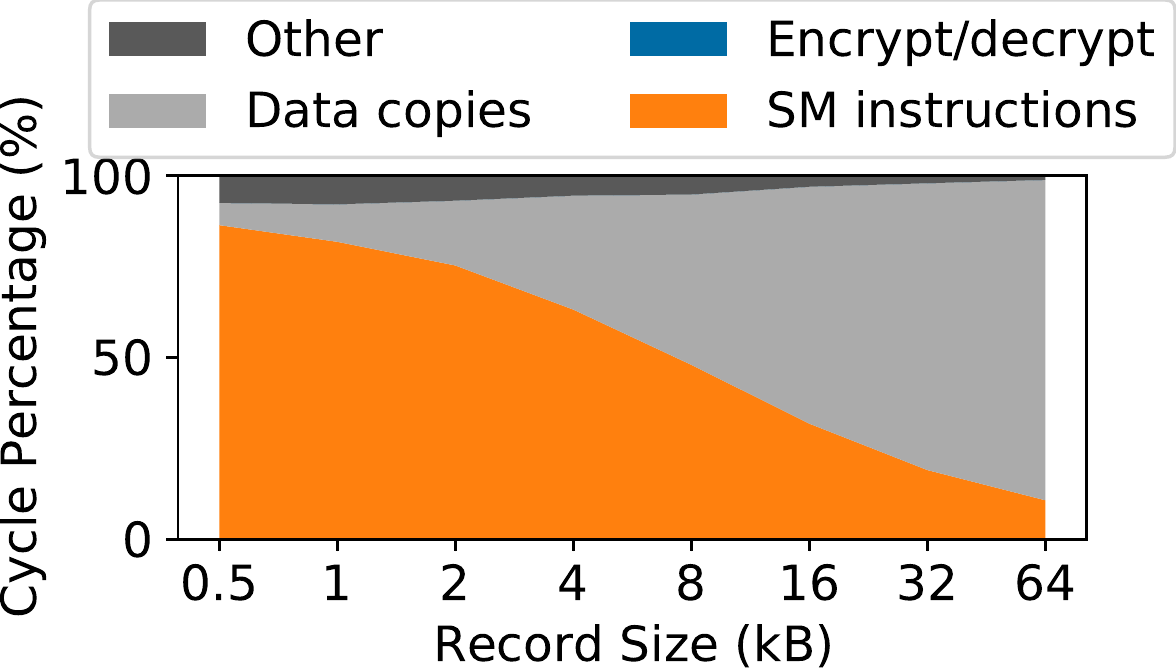}
}
\subfloat[\stspatial]{
\includegraphics[width=0.27\linewidth]{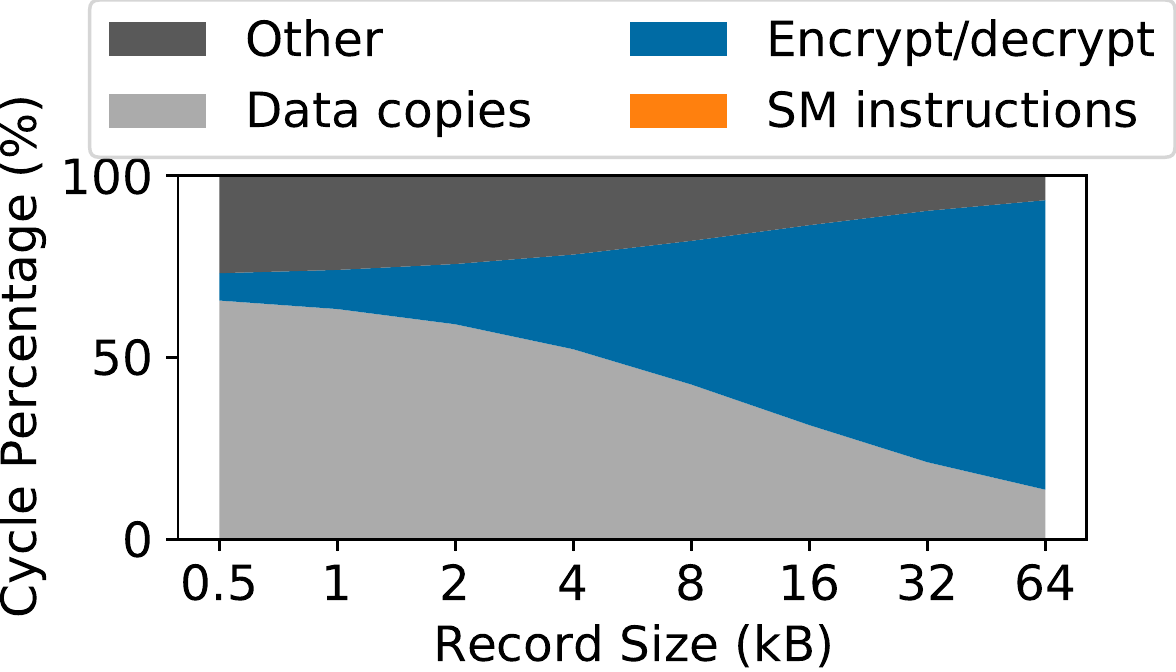}
}
\caption{Performance breakdown for the proxy pattern.}
\label{fig:icall_breakdown}
\end{figure*}

\paragraph{Synthetic Benchmark: Data-Sharing Patterns.}
We construct synthetic benchmarks for 
the $3$ patterns in Section~\ref{sec:baseline-model} and
measure data sharing overhead (excluding any actual data processing). We set up $2$ (for producer-consumer and client-server)
or $3$ (for the proxy pattern) enclaves and compare: (a)~full
\codename support as described in Section~\ref{sec:design}
(\codename-full); (b)~\codename without the lock permission bit design
(\codename-nolock); and (c)~spatial isolation which transfers data
through secure public memory. Figure~\ref{fig:icall_results} shows the
performance for $3$ patterns. Figure~\ref{fig:icall_breakdown} shows the 
breakdown for the proxy pattern.

\emph{Observations:} 
The results exhibit a huge performance improvement of \stcodename-full
over \stspatial, which increases with an increase in the record size.
When the record size is $512$ bytes, \stcodename-full provides over
$60\times$ speedup compared with \stspatial; when the record size
increases to $64$KB the speedup also increases and reaches
$600\times$. In \stcodename-full, although invoking security
instructions is a large contributor to the overhead, by doing this the
application eliminates copying and communication through secure public
memory. As a result, the total overhead of \stcodename-full does not
increase with the size of the transferred data, unlike \stspatial.
Note that \stcodename-full corresponds to a two-way isolation paradigm
highlighted in Section~\ref{sec:pattern-costs}.

\stcodename-nolock, the design of \codename with the lock permission
bit removed, is shown to be more costly than \stcodename-full with
overhead that increases with the data size.
Figure~\ref{fig:icall_breakdown} indicates that this is because
\stcodename-nolock does not completely eliminate data copying. 

\paragraph{Synthetic Benchmark: Thread Synchronization.}
We implement a common workload for spinlocks between threads, each
of which runs in a separate enclave, but both do not trust the OS. 
For \codename, we further distinguish simple spinlocks
(\codename-spinlock) and \futex{}es (\codename-futex). For spinlocks,
we keep the lock state in a shared region with no access to the OS.
For futexes, the untrusted OS has read-only access to the lock states,
which allows enclaves to sleep while waiting for locks and be woken up
by the OS when locks are released. This form of sharing corresponds
the one-way isolation described in Section~\ref{sec:pattern-costs},
where the OS has read-only permissions.
For \stspatial, we implement a dedicated trusted
coordinator enclave to manage the lock states, with 
enclaves communicating with it through secure public memory for lock
acquisition and release. 

\begin{figure}[!tb]
\centering
\includegraphics[width=0.75\linewidth]{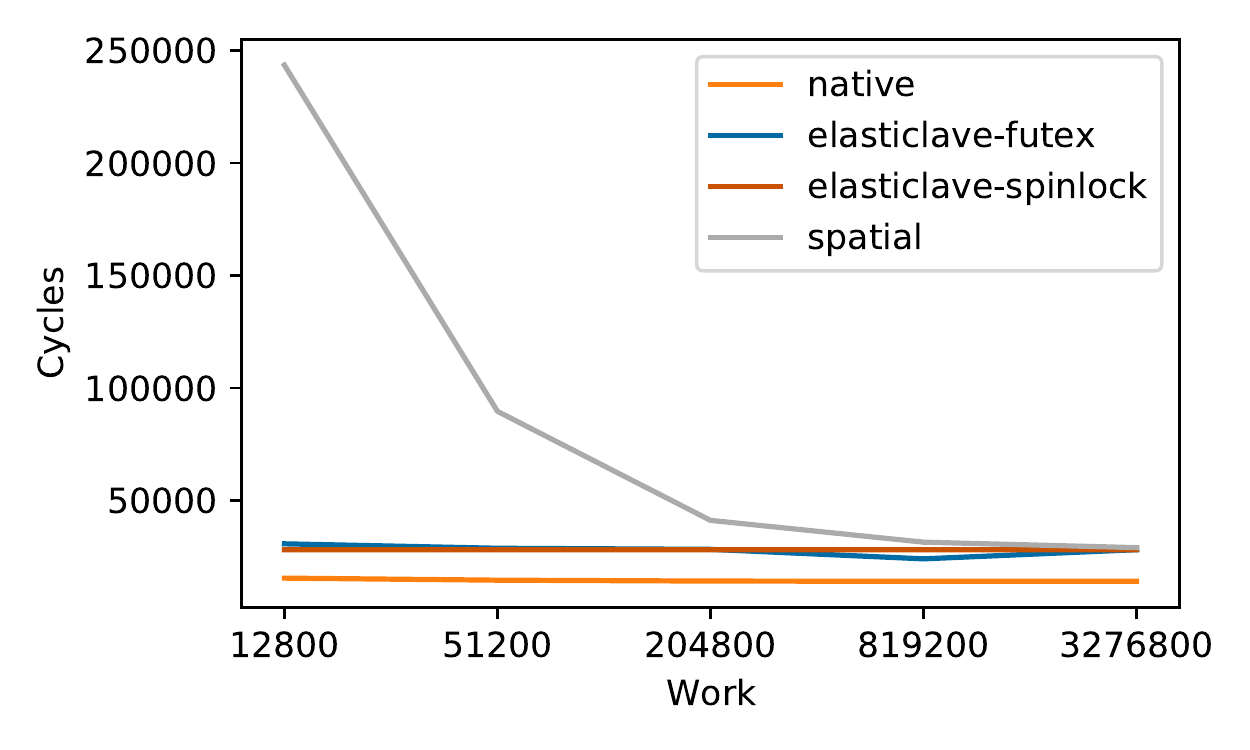}
\caption{Synthetic Thread Synchronization Performance. Cycles
(normalized by the contention).} 
\label{fig:lock-basic}
\end{figure}

\begin{figure}[!tb]
    \centering
    \includegraphics[width=0.7\linewidth]{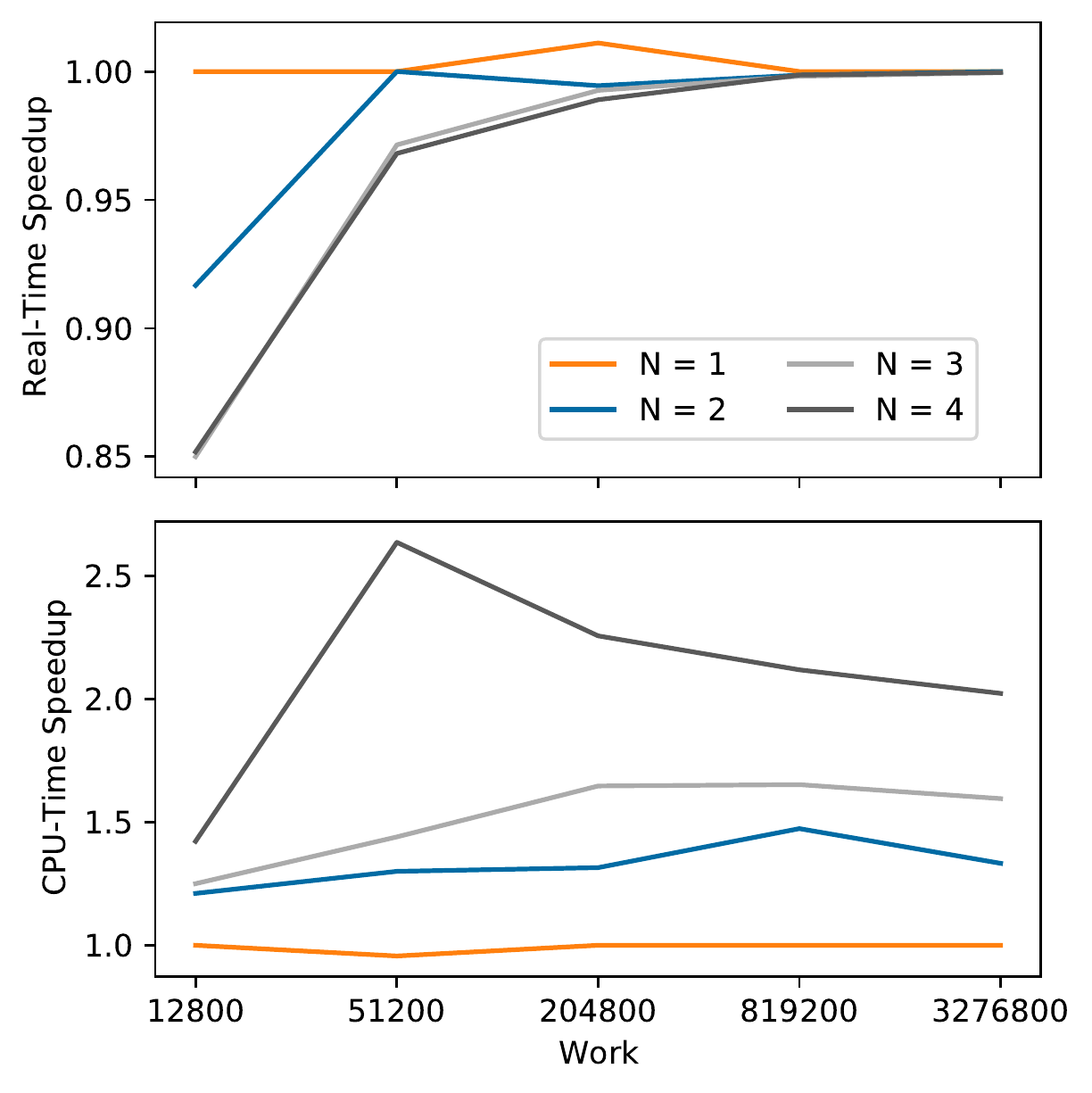}
    \caption{
    \stcodename-futex vs \stcodename-spinlock.} 
    \label{fig:lock-futex}
\end{figure}

\emph{Observations:} 
We report that \codename-futex and \codename-spinlock
achieve much higher performance compared with \stspatial (Figures~\ref{fig:lock-basic}), especially
when the contention is low (the lock is acquired and released often).
For higher contention where the time spent waiting for
the lock overshadows the overhead of acquiring and releasing the lock,
the $3$ settings have comparable performance. 
In addition,
\codename-futex achieves up to $1.5\times$ CPU-time performance
improvement over \codename-spinlock despite having no advantage in
terms of real-time performance (wall-clock latency). 
Figure~\ref{fig:lock-futex} shows the performance of 
\stcodename-futex vs \stcodename-spinlock.

\begin{figure*}[!t] \centering
\begin{minipage}[b]{0.5\linewidth}
\subfloat[IOZone Writer \label{subfig:iozone-writer-native}]{
\includegraphics[width=.45\linewidth]{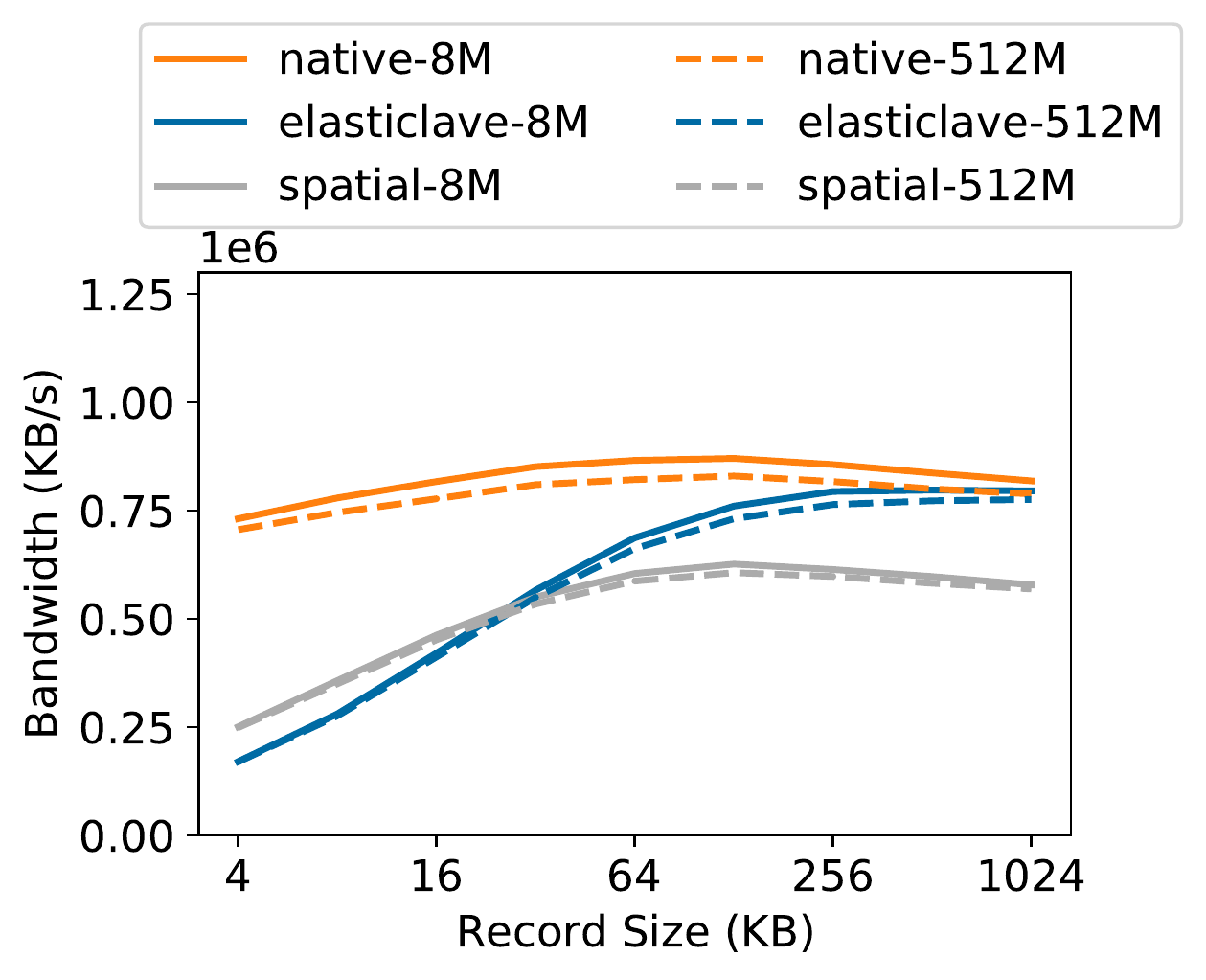}
}
\subfloat[IOZone Reader \label{subfig:iozone-reader-native}]{
\includegraphics[width=.45\linewidth]{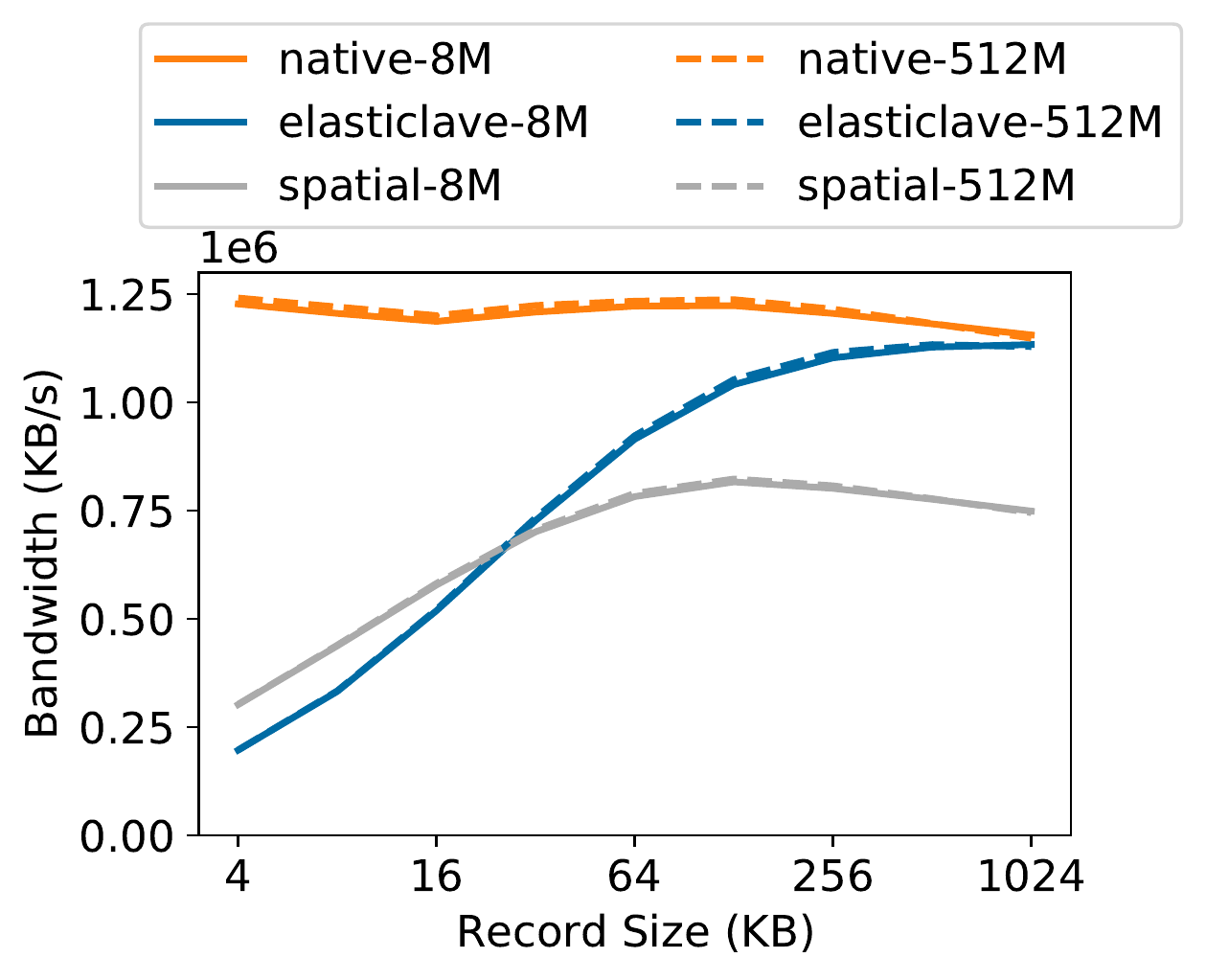}
}
\captionof{figure}{IOZone Bandwidth for 8M and 512M byte files.}
\label{fig:iozone}
\end{minipage}
\hfill
\begin{minipage}[b]{0.45\linewidth}
{
\includegraphics[width=0.95\linewidth]{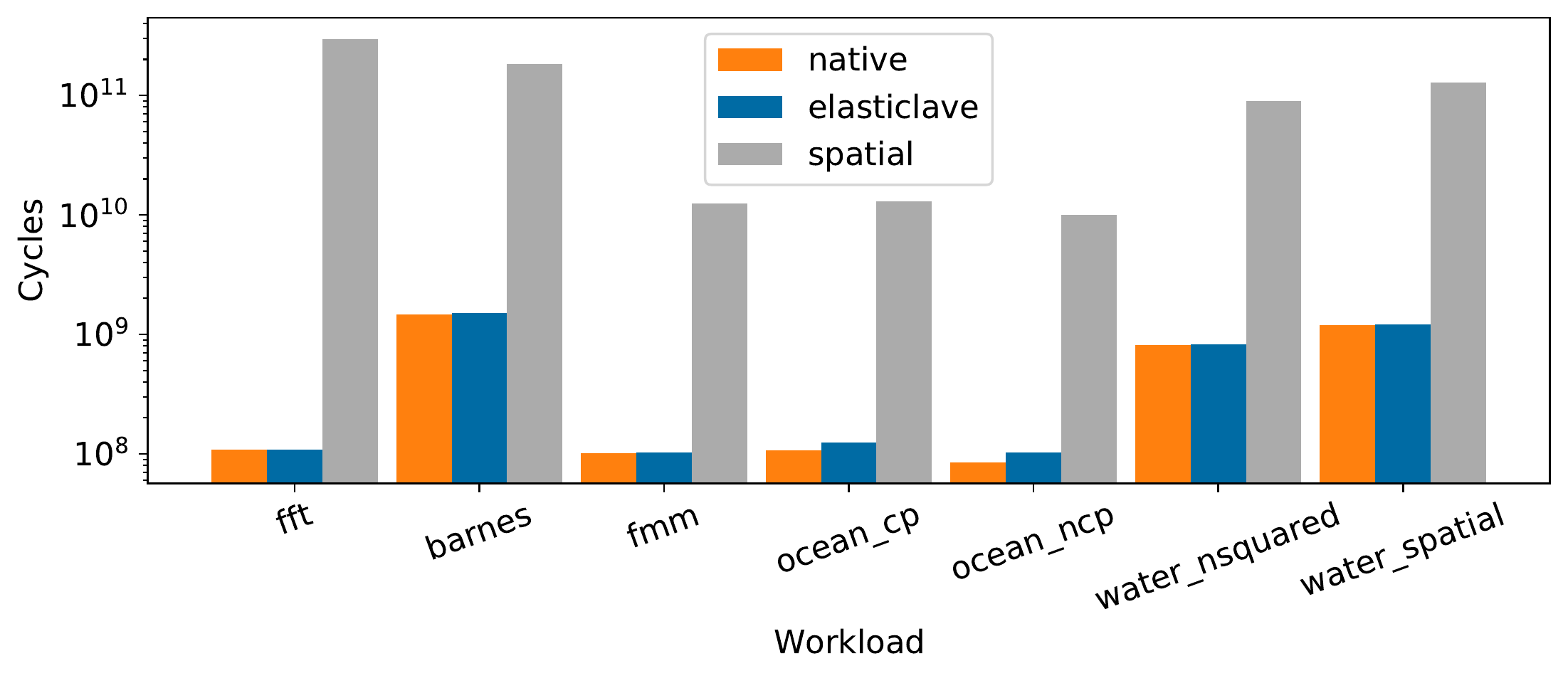}
}
\captionof{figure}{SPLASH-2 wall-clock time, measured in cycles.}
\label{fig:splash2}
\end{minipage}
\end{figure*}

\paragraph{Real-World Benchmark 1: File I/O.}
We run the IOZone benchmark ~\cite{iozone}; it makes frequent file I/O
calls from enclave into the untrusted host process. Here, for \stspatial, the communication
does not need to be protected with secure public memory. Figures~\ref{subfig:iozone-writer-native} and ~\ref{subfig:iozone-reader-native}
shows write and read bandwidth.

\emph{Observations:} 
Even without secure public memory
communication in \stspatial, 
\codename achieves a higher bandwidth than \stspatial, when the record
size grows above a threshold ($16$KB). The bandwidth increase reaches
as high as $0.4$ for the writer workload and around $0.5$ for the
reader workload when the record size is sufficiently large.

\paragraph{Real-World Benchmark 2: Parallel Computation.}
We ran $7$ SPLASH-2 workloads in a two-enclave setting. We adapted the
workloads to multi-enclave implementations by collecting together the
data shared across threads in a memory region which would be shared
across the enclaves. For \stspatial, load/store instructions that
operate on the memory region are trapped and emulated with RPCs by the
enclave runtime. 
Figure~\ref{fig:splash2} shows numbers of cycles to execute parallel workloads in two
enclaves (excluding initialization).
We were not able to run 
$\tt{libsodium}$ inside the enclave runtime, so we did not use
encryption-decryption when copying data to-and-from secure
public memory for \stspatial in this experiment. 
So the actual  overhead in a secure implementation would be
higher than reported in  here. Thus, even if the
processor had support for cryptographic accelerators (e.g., AES-NI)
that may speedup \stspatial, \codename speedups just due to
zero-copies are still significant and out-perform \stspatial.

\emph{Observations:} 
On all the workloads measured, \stcodename is $2$-$3$ orders of
magnitude faster than \stspatial.

\begin{figure}[!tb]
    \centering
    \includegraphics[width=.75\linewidth]{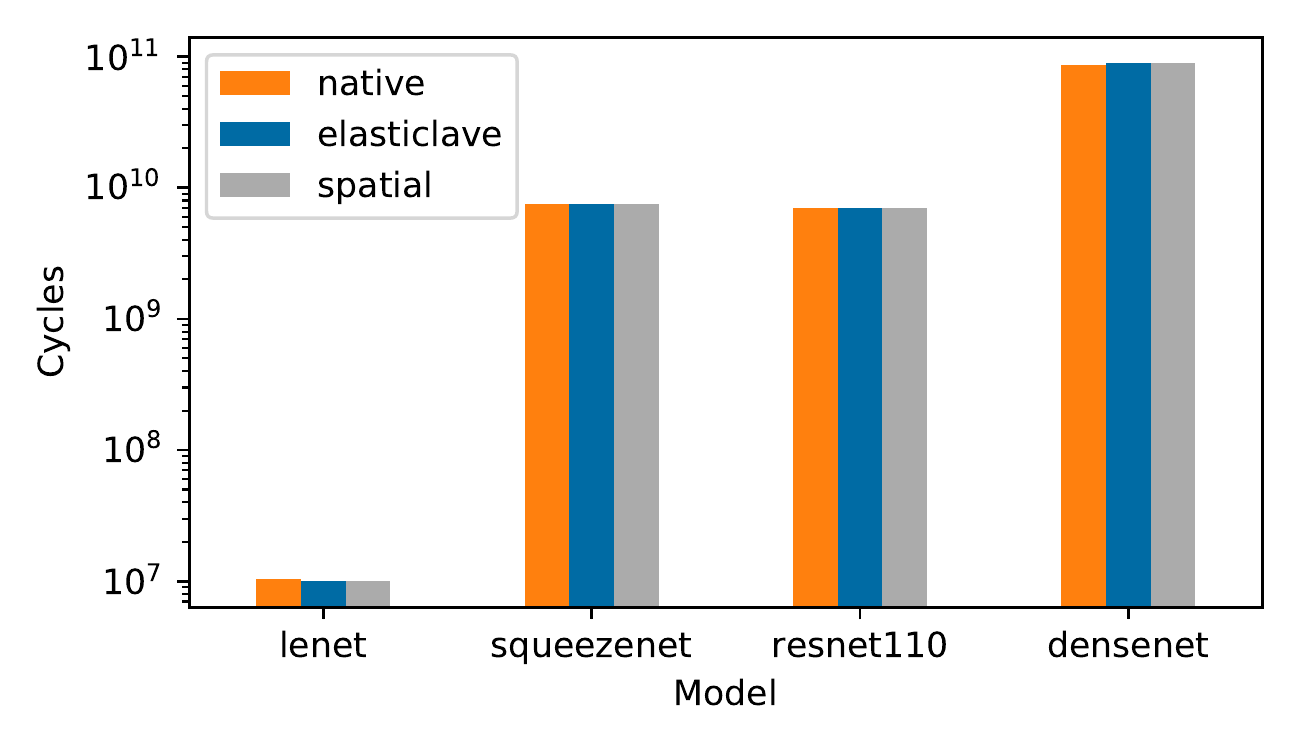}
    \caption{Cycles spent running each ML model.}
    \label{fig:privado}
\end{figure}

\paragraph{Real-World Benchmark 3: ML Inference.}
We run $4$ machine learning models for image
classification~\cite{privado} to measure \codename performance on applications with minimal
data sharing needs (Figure~\ref{fig:privado}).
Each of the inference models runs with a single enclave thread 
and one shared region between
the enclave and the OS to loads input images.

\emph{Observations:} 
The $3$ settings have similar performance. Thus,  \codename does not
slow-down CPU-intensive applications that do not share data
extensively.

\begin{framed}
Compared to \stspatial,  \codename improves I/O-intensive 
workload performance up to 
$600\times$ (data size $ > 64$KB) and demonstrates $50$\% higher 
bandwidth. For shared-memory benchmarks, it gains  
up to a $1000\times$ speedup. 
\end{framed}

\paragraph{Comparison to Native.}
 \codename 
is up to $90\%$ as performant as \stnative (traditional Linux
processes no enclave isolation) for a range of our benchmarks (Figures~\ref{fig:splash2} and ~\ref{fig:privado}). 
\codename performs comparable to \stnative for frequent data
sharing over large-size (Figures~\ref{subfig:iozone-writer-native} and ~\ref{subfig:iozone-reader-native}).

\subsection{Impact on Implementation Complexity}
\label{sec:overhead}
We report on the \codename TCB, hardware chip area, context switch cost, and critical path latency.

\paragraph{TCB.}
\codename does not incur any change to the hardware. Its only
requires additional PMP entries, one entry per shared region. 
\codename TCB is $6814$ \loc (Table~\ref{table:tcb}). $3085$ \loc 
to implement the \codename interface in $\tt{m-mode}$, $3729$ \loc to use the interface in an enclave.

\begin{table}[!tb]
\centering
\resizebox{0.4\textwidth}{!}{
\begin{tabular}{lrr}
\toprule
    \textbf{Function} & \textbf{\codename} & \textbf{Enclave} \\
                      & \textbf{Privileged TCB} & \textbf{Runtime} \\
\midrule
    \uid management & 1070 & 0 \\
    Permission matrix enforcement & 574 & 0 \\
    \codename instruction interface & 219 & 82 \\
    Argument marshaling & 0 & 88 \\
    Wrappers for \codename interface & 0 & 1407 \\
    Miscellaneous & 960 & 1869 \\

\midrule
    Total & 3085 & 3729 \\
\bottomrule
\end{tabular}
}

\caption{Breakdown in LoC of \codename TCB \& Enclave runtime libraries (not in \codename TCB).}
\label{table:tcb}
\end{table} 
\paragraph{Context Switches.}
Context switching between enclaves and the OS 
incurs PMP changes. 
Thus, the overhead may change with
numbers of PMP-protected memory regions. To empirically measure this, 
we record the percentage of cycles spent on
context switches in either direction for a workload that
never explicitly switches out to the OS (therefore all context
switches from the enclave to the OS are due to interrupts).
The percentage overhead increases 
linearly with the number of memory regions but is negligibly
small: $0.1\%$ for $1$ memory region and $0.15\%$ for $4$ memory
regions. 

\newcommand{\delayfreq}{$1$ GHz}

\paragraph{Hardware Critical Path Delay.} 
To determine the
critical path of the hardware design
and examine if the PMP entries are on this path,
we push the design to a target frequency of
\delayfreq{}\footnote{We set the frequency higher than
\targetfreq{} (which is what we have for our successful synthesis) to
push the optimization limit of the hardware design so we can
find out the bottleneck of the hardware design.}.  We measure
the global critical path latency, which is of the whole core, and the
critical path through the PMP registers. 
With this, we compute the slack, which is the desired delay minus the actual delay---a larger slack corresponds to a smaller actual delay in comparison to the desired delay. We find that the slack through PMP is significantly better than global critical path. With $16$ PMP entries, the slack through PMP is $-44.1$ picoseconds compared to $-190.1$ picoseconds for the global critical path. 
In other words, the PMP would allow for a higher clock speed, but the rest of the design prevents it.
Thus, the number of PMP entries is {\em not the bottleneck of the
timing of the hardware design}.
We also tested that PMPs are not on the critical path for 8 and 32 PMP settings as well (details elided due to space). As a direct result, the number of PMP entries does not create a performance bottleneck for any instruction (e.g., load/store, PMP read/write), in our tests.

\begin{figure}[!tb]
    \centering
    \includegraphics[width=0.6\linewidth]{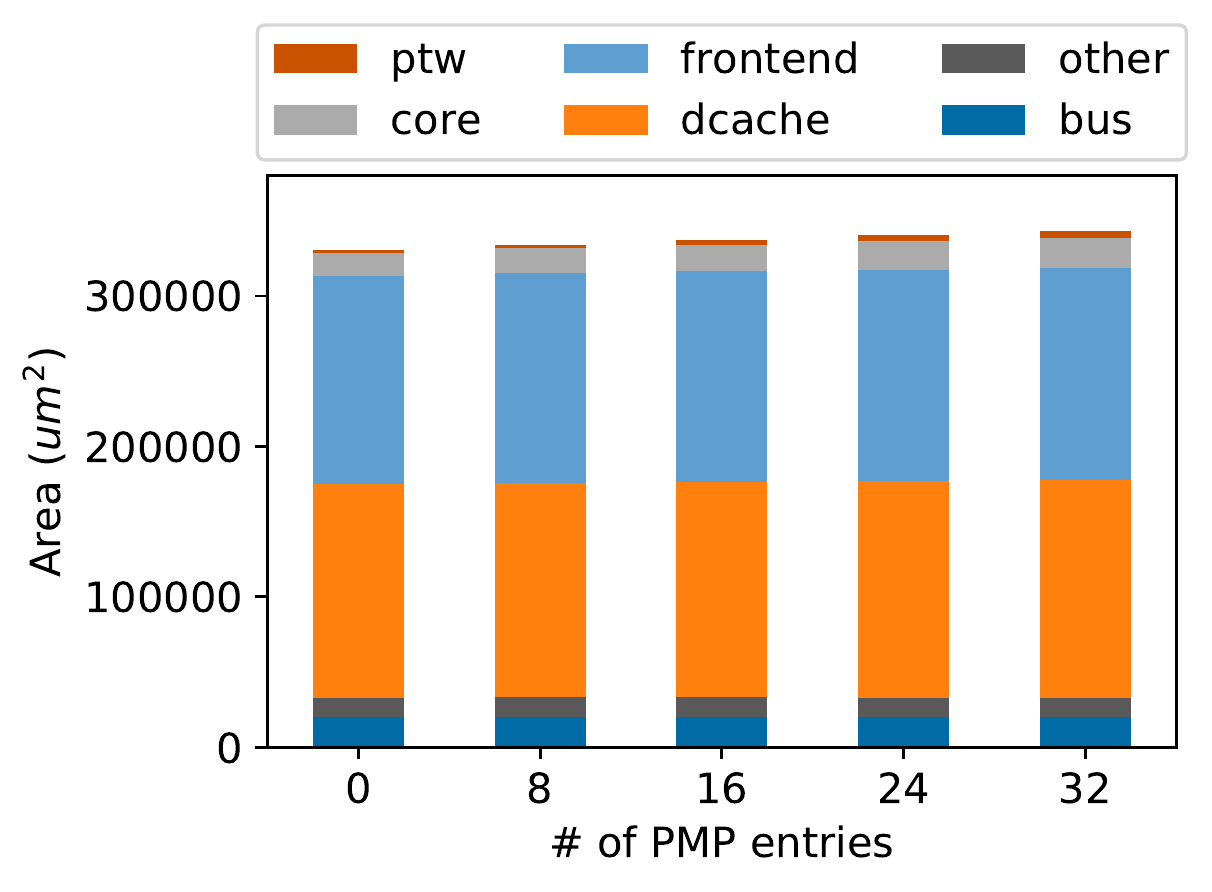}
     \caption{ RocketChip Area vs. numbers of PMP entries.}
     \label{fig:area-breakdown}
\end{figure}

\paragraph{Area.}
The only impact of \codename on \riscv hardware is on the increased
PMP pressure (i.e., we require more PMP entries per region), which
increases chip area requirements. We synthesize RocketChip with
different numbers of PMP registers and collect the area statistics.
The range we explore goes beyond the limit of $16$ in the standard
\riscv ISA specification. Figure~\ref{fig:area-breakdown}
exhibits the increase in the total area with increasing numbers of PMP
entries. The increase is not significant. Starting with $0$ PMP
entries, every $8$ additional PMP entries only incurr $1\%$
increase in the total area. 

\begin{framed}
\codename does not significantly increases software TCB size
(\textasciitilde{}$6800$ LOC), critical path delay, or
hardware area pressure
(\textasciitilde{}$1\%$ per 8 PMP entries), which shows that
the design scales well with number of regions.

\end{framed}
 \section{Related Work}
\label{sec:related}

Isolation abstractions are of long-standing importance to security.
There has been extensive work on partitioning security-critical
applications using software isolation abstractions using namespace
isolation (e.g., containers), software-based techniques (e.g., 
SFI~\cite{sfi}, native-client~\cite{nacl}), language-based isolation
(java-capabilities~\cite{mettler2010joe},
web-sandboxing~\cite{jit-sandboxing}), OS-based
sandboxing~\cite{tx-box}, and using hypervisors (e.g., virtual
machines~\cite{xen, sp3}). Our work is on hardware support for
isolation, namely using enclave TEEs.
\codename draws attention to a single point in the design space of
memory models that TEEs support, specifically, its memory model and
its impact on memory sharing. The pre-dominant model used today is
that of spatial isolation, which is used in Intel SGX, as well as
others (e.g., TrustZone~\cite{armtrustzone}, AMD SEV~\cite{amd-sev,
amd-sev-es}). \codename explains the conceptual drawbacks of this
model and offers a relaxation that enables better performance. Intel
SGX v2 follows the spatial isolation design, with the exception that
permissions and sizes of private regions can be changed
dynamically~\cite{mckeen2016sgx, xing2016sgx}. As a result, the
"all-or-none" trust division between enclaves remains the same as
in v1.

TEEs, including SGX, has received extensive security scrutiny. Recent
works have shown the challenges of leakage via side-channels. New TEEs
have focused on providing better micro-architectural resistance to
side-channels~\cite{mi6}. Designs providing stronger confidentiality
due to oblivious execution techniques have been
proposed~\cite{maas2013phantom}. Keystone is a framework for
experimenting with new enclave designs, which uses the spatial
isolation model~\cite{dayeol2020keystone}. 

Several TEE designs have been proposed prior to Intel SGX which show
the promise and feasibility of enclave
TEEs~\cite{bastion,shinde2015podarch, secureme,
secureblue, xom, aegis}. 
After the availability of SGX in commercial
CPUs~\cite{sgx2014progref}, several works have shed light on the
security and compatibility improvements possible over the original SGX
design~\cite{costan2016sanctum,dayeol2020keystone,ferraiuolo2017komodo,mckeen2016sgx,xing2016sgx,park2020nestedenclave,mi6,sanctuary}.
They allow for better security, 
additional classes of applications, better memory allocation,
and  hierarchical security protection.
Nevertheless, they largely adhere to the {\em spatial isolation
model}, and have not explicitly challenged the assumptions.

TEE-based designs for memory sharing and TCB reduction are similar in
spirit to mechanisms used in hypervisors and microkernels---for
example, as used in page-sharing via EPT tables~\cite{cloudvisor,
chaos}, IOMMU implementations for memory-mapped devices such as GPUs
or NICs~\cite{damn}.
The key difference is in the trust model: Hypervisors~\cite{xen} and
microkernels~\cite{liedtke1995microkernel} are entrusted to make
security decisions on behalf of VMs, whereas in enclaved
TEEs, the privileged software is untrusted and enclaves self-arbitrate
security decisions. Further, microkernels and monolithic kernels
operate in system mode (e.g., S-mode in \riscv) which is in the
TCB. They are larger compared to (say) the \codename TCB.

Emerging proposals such as Intel TDX~\cite{tdx}, Intel MKTME~\cite{mktme},
Intel MPK~\cite{mpk}, Donky~\cite{donky} enable hardware-enforced domain protection.
However, they protect entire virtual machines or
groups of memory pages (in contrast to enlcaves in Intel
SGX). 
Notably, they extend fast hardware support to protect physical memory
of a trust domain (e.g., from a physical adversary) but 
adhere to spatial model. They can benefit from \codename memory model. \section{Conclusion}

We present \codename, a new TEE memory model that allows enclaves to
selectively and temporarily share memory with other enclaves and the
OS. 
We demonstrate that \codename eliminates expensive data copy operation 
and maintains same level of application-desired security.
Our \codename prototype on RISC-V FPGA core offers $1$ to $2$ order of
magnitude performance improvements over existing models.
 
\section*{Acknowledgments}
We thank 
Aashish Kolluri, 
Burin Amornpaisannon, 
Dawn Song, 
Dayeol Lee, 
Jialin Li, 
Li-Shiuan Peh, 
Roland Yap, 
Shiqi Shen, 
Yaswanth Tavva, 
Yun Chen, and 
Zhenkai Liang.
for their feedback on improving earlier drafts of the paper.
The authors acknowledge the support from the Crystal Center and 
Singapore National Research Foundation ("SOCure`` grant
NRF2018NCR-NCR002-0001 www.green-ic.org/socure).
This material is in part based upon work supported by the National
Science Foundation under Grant No. DARPA N66001-15-C-4066 and Center
for Long-Term Cybersecurity. Any opinions, findings, and conclusions
or recommendations expressed in this material are those of the
authors and do not necessarily reflect the views of the National
Science Foundation.  
\bibliographystyle{plain}

\begin{thebibliography}{10}

\bibitem{jit-sandboxing}
Jason Ansel, Petr Marchenko, {\'U}lfar Erlingsson, Elijah Taylor, Brad Chen,
  Derek~L Schuff, David Sehr, Cliff~L Biffle, and Bennet Yee.
\newblock Language-independent sandboxing of just-in-time compilation and
  self-modifying code.
\newblock In {\em Proceedings of the 32nd ACM SIGPLAN conference on Programming
  language design and implementation}, pages 355--366, 2011.

\bibitem{armtrustzone}
Arm trustzone technology.
\newblock \url{https://developer.arm.com/ip-products/security-ip/trustzone}.

\bibitem{scone}
Sergei Arnautov, Bohdan Trach, Franz Gregor, Thomas Knauth, Andr{\'{e}} Martin,
  Christian Priebe, Joshua Lind, Divya Muthukumaran, Dan O'Keeffe, Mark
  Stillwell, David Goltzsche, David~M. Eyers, R{\"{u}}diger Kapitza, Peter~R.
  Pietzuch, and Christof Fetzer.
\newblock {SCONE:} secure linux containers with intel {SGX}.
\newblock In {\em {OSDI}}, pages 689--703. {USENIX} Association, 2016.

\bibitem{rocket}
Krste Asanovic, Rimas Avizienis, Jonathan Bachrach, Scott Beamer, David
  Biancolin, Christopher Celio, Henry Cook, Daniel Dabbelt, John Hauser, Adam
  Izraelevitz, et~al.
\newblock The rocket chip generator.
\newblock {\em EECS Department, University of California, Berkeley, Tech. Rep.
  UCB/EECS-2016-17}, 2016.

\bibitem{xen}
Paul Barham, Boris Dragovic, Keir Fraser, Steven Hand, Tim Harris, Alex Ho,
  Rolf Neugebauer, Ian Pratt, and Andrew Warfield.
\newblock Xen and the art of virtualization.
\newblock In {\em SOSP}, 2003.

\bibitem{haven}
Andrew Baumann, Marcus Peinado, and Galen~C. Hunt.
\newblock Shielding applications from an untrusted cloud with haven.
\newblock In {\em {OSDI}}, pages 267--283. {USENIX} Association, 2014.

\bibitem{splash-summary}
Christian Bienia.
\newblock {\em Benchmarking Modern Multiprocessors}.
\newblock PhD thesis, Princeton University, January 2011.

\bibitem{secureblue}
Rick Boivie and Peter Williams.
\newblock Secureblue++: Cpu support for secure execution.
\newblock {\em IBM, IBM Research Division, RC25287 (WAT1205-070)}, pages 1--9,
  2012.

\bibitem{mi6}
Thomas Bourgeat, Ilia~A. Lebedev, Andrew Wright, Sizhuo Zhang, Arvind, and
  Srinivas Devadas.
\newblock {MI6:} secure enclaves in a speculative out-of-order processor.
\newblock In {\em {MICRO}}, pages 42--56. {ACM}, 2019.

\bibitem{sanctuary}
Ferdinand Brasser, David Gens, Patrick Jauernig, Ahmad-Reza Sadeghi, and
  Emmanuel Stapf.
\newblock Sanctuary: Arming trustzone with user-space enclaves.
\newblock In {\em NDSS}, 2019.

\bibitem{bastion}
D.~{Champagne} and R.~B. {Lee}.
\newblock Scalable architectural support for trusted software.
\newblock In {\em HPCA - 16 2010 The Sixteenth International Symposium on
  High-Performance Computer Architecture}, pages 1--12, 2010.

\bibitem{chaos}
Haibo Chen, Fengzhe Zhang, Cheng Chen, Ziye Yang, Rong Chen, Binyu Zang, and
  Wenbo Mao.
\newblock {Tamper-Resistant Execution in an Untrusted Operating System Using A
  Virtual Machine Monitor}, 2007.

\bibitem{overshadow}
Xiaoxin Chen, Tal Garfinkel, E.~Christopher Lewis, Pratap Subrahmanyam, Carl~A.
  Waldspurger, Dan Boneh, Jeffrey Dwoskin, and Dan~R.K. Ports.
\newblock Overshadow: A virtualization-based approach to retrofitting
  protection in commodity operating systems.
\newblock In {\em Proceedings of the 13th International Conference on
  Architectural Support for Programming Languages and Operating Systems},
  ASPLOS XIII, page 2–13, New York, NY, USA, 2008. Association for Computing
  Machinery.

\bibitem{secureme}
Siddhartha Chhabra, Brian Rogers, Yan Solihin, and Milos Prvulovic.
\newblock {SecureME: A Hardware-software Approach to Full System Security}.
\newblock In {\em ICS}, 2011.

\bibitem{sgxexplained}
Victor Costan and Srinivas Devadas.
\newblock Intel {SGX} explained.
\newblock {\em {IACR} Cryptol. ePrint Arch.}, 2016:86, 2016.

\bibitem{costan2016sanctum}
Victor Costan, Ilia Lebedev, and Srinivas Devadas.
\newblock Sanctum: Minimal hardware extensions for strong software isolation.
\newblock In {\em 25th {USENIX} Security Symposium ({USENIX} Security 16)},
  pages 857--874, Austin, TX, August 2016. {USENIX} Association.

\bibitem{m2r}
Tien Tuan~Anh Dinh, Prateek Saxena, Ee{-}Chien Chang, Beng~Chin Ooi, and
  Chunwang Zhang.
\newblock {M2R:} enabling stronger privacy in mapreduce computation.
\newblock In {\em {USENIX} Security Symposium}, pages 447--462. {USENIX}
  Association, 2015.

\bibitem{ferraiuolo2017komodo}
Andrew Ferraiuolo, Andrew Baumann, Chris Hawblitzel, and Bryan Parno.
\newblock Komodo: Using verification to disentangle secure-enclave hardware
  from software.
\newblock In {\em Proceedings of the 26th Symposium on Operating Systems
  Principles}, SOSP ’17, page 287–305, New York, NY, USA, 2017. Association
  for Computing Machinery.

\bibitem{vif}
D.~{Gong}, M.~{Tran}, S.~{Shinde}, H.~{Jin}, V.~{Sekar}, P.~{Saxena}, and M.~S.
  {Kang}.
\newblock Practical verifiable in-network filtering for ddos defense.
\newblock In {\em 2019 IEEE 39th International Conference on Distributed
  Computing Systems (ICDCS)}, 2019.

\bibitem{sgx-box}
Juhyeng Han, Seongmin Kim, Jaehyeong Ha, and Dongsu Han.
\newblock Sgx-box: Enabling visibility on encrypted traffic using a secure
  middlebox module.
\newblock In {\em Proceedings of the First Asia-Pacific Workshop on
  Networking}, APNet'17, 2017.

\bibitem{iozone}
Iozone filesystem benchmark.
\newblock \url{http://iozone.org}.

\bibitem{tx-box}
S.~{Jana}, D.~E. {Porter}, and V.~{Shmatikov}.
\newblock Txbox: Building secure, efficient sandboxes with system transactions.
\newblock In {\em 2011 IEEE Symposium on Security and Privacy}, pages 329--344,
  2011.

\bibitem{amd-sev-es}
David Kaplan.
\newblock {AMD SEV-ES}.
\newblock
  \url{http://support.amd.com/TechDocs/ProtectingVMRegisterStatewithSEV-ES.pdf},
  2017.

\bibitem{amd-sev}
David Kaplan, Jeremy Powell, and Tom Woller, 2016.

\bibitem{DBLP:conf/isca/KarandikarMKBAL18}
Sagar Karandikar, Howard Mao, Donggyu Kim, David Biancolin, Alon Amid, Dayeol
  Lee, Nathan Pemberton, Emmanuel Amaro, Colin Schmidt, Aditya Chopra, Qijing
  Huang, Kyle Kovacs, Borivoje Nikolic, Randy~H. Katz, Jonathan Bachrach, and
  Krste Asanovic.
\newblock Firesim: Fpga-accelerated cycle-exact scale-out system simulation in
  the public cloud.
\newblock In {\em {ISCA}}, pages 29--42. {IEEE} Computer Society, 2018.

\bibitem{dayeol2020keystone}
Dayeol Lee, David Kohlbrenner, Shweta Shinde, Krste Asanovi\'{c}, and Dawn
  Song.
\newblock Keystone: An open framework for architecting trusted execution
  environments.
\newblock In {\em Proceedings of the Fifteenth European Conference on Computer
  Systems}, New York, NY, USA, 2020. Association for Computing Machinery.

\bibitem{xom}
David Lie, Chandramohan~A. Thekkath, Mark Mitchell, Patrick Lincoln, Dan Boneh,
  John~C. Mitchell, and Mark Horowitz.
\newblock Architectural support for copy and tamper resistant software.
\newblock In {\em {ASPLOS}}, pages 168--177. {ACM} Press, 2000.

\bibitem{liedtke1995microkernel}
Jochen Liedtke.
\newblock On micro-kernel construction.
\newblock {\em ACM SIGOPS Operating Systems Review}, 29(5):237--250, 1995.

\bibitem{maas2013phantom}
Martin Maas, Eric Love, Emil Stefanov, Mohit Tiwari, Elaine Shi, Krste
  Asanovic, John Kubiatowicz, and Dawn Song.
\newblock {PHANTOM:} practical oblivious computation in a secure processor.
\newblock In {\em {ACM} Conference on Computer and Communications Security},
  pages 311--324. {ACM}, 2013.

\bibitem{damn}
Alex Markuze, Igor Smolyar, Adam Morrison, and Dan Tsafrir.
\newblock {DAMN:} overhead-free {IOMMU} protection for networking.
\newblock In {\em {ASPLOS}}, pages 301--315. {ACM}, 2018.

\bibitem{rote}
Sinisa Matetic, Mansoor Ahmed, Kari Kostiainen, Aritra Dhar, David Sommer,
  Arthur Gervais, Ari Juels, and Srdjan Capkun.
\newblock {ROTE}: Rollback protection for trusted execution.
\newblock In {\em 26th {USENIX} Security Symposium ({USENIX} Security 17)},
  2017.

\bibitem{mckeen2016sgx}
Frank McKeen, Ilya Alexandrovich, Ittai Anati, Dror Caspi, Simon Johnson,
  Rebekah Leslie-Hurd, and Carlos Rozas.
\newblock Intel® software guard extensions (intel® sgx) support for dynamic
  memory management inside an enclave.
\newblock In {\em Proceedings of the Hardware and Architectural Support for
  Security and Privacy 2016}, HASP 2016, New York, NY, USA, 2016. Association
  for Computing Machinery.

\bibitem{sgx}
Frank McKeen, Ilya Alexandrovich, Alex Berenzon, Carlos~V. Rozas, Hisham Shafi,
  Vedvyas Shanbhogue, and Uday~R. Savagaonkar.
\newblock Innovative instructions and software model for isolated execution.
\newblock In {\em HASP@ISCA}, page~10. {ACM}, 2013.

\bibitem{mettler2010joe}
Adrian Mettler, David~A Wagner, and Tyler Close.
\newblock Joe-e: A security-oriented subset of java.
\newblock In {\em NDSS}, volume~10, pages 357--374, 2010.

\bibitem{mktme}
Intel releases new technology specification for memory encryption.
\newblock
  \url{https://software.intel.com/content/www/us/en/develop/blogs/intel-releases-new-technology-specification-for-memory-encryption.html?wapkw=tme}.

\bibitem{mpk}
Intel® 64 and ia-32 architectures software developer manual, 2018.

\bibitem{park2020nestedenclave}
J.~{Park}, N.~{Kang}, T.~{Kim}, Y.~{Kwon}, and J.~{Huh}.
\newblock Nested enclave: Supporting fine-grained hierarchical isolation with
  sgx.
\newblock In {\em 2020 ACM/IEEE 47th Annual International Symposium on Computer
  Architecture (ISCA)}, pages 776--789, 2020.

\bibitem{riscvpriv}
The risc-v instruction set manual: Volume ii: Privileged architecture.
\newblock \url{https://riscv.org/specifications/privileged-isa}.

\bibitem{donky}
David Schrammel, Samuel Weiser, Stefan Steinegger, Martin Schwarzl, Michael
  Schwarz, Stefan Mangard, and Daniel Gruss.
\newblock Donky: Domain keys {\textendash} efficient in-process isolation for
  risc-v and x86.
\newblock In {\em 29th {USENIX} Security Symposium ({USENIX} Security 20)},
  pages 1677--1694. {USENIX} Association, August 2020.

\bibitem{vc3}
Felix Schuster, Manuel Costa, Cedric Fournet, Christos Gkantsidis, Marcus
  Peinado, Gloria Mainar-Ruiz, and Mark Russinovich.
\newblock {VC3: Trustworthy Data Analytics in the Cloud}.
\newblock In {\em IEEE S\&P}, 2015.

\bibitem{sgx2014progref}
Software guard extensions programming reference rev. 2.
\newblock \url{http://software.intel.com/sites/default/files/329298-002.pdf},
  October 2014.

\bibitem{occlum}
Youren Shen, Hongliang Tian, Yu~Chen, Kang Chen, Runji Wang, Yi~Xu, Yubin Xia,
  and Shoumeng Yan.
\newblock Occlum: Secure and efficient multitasking inside a single enclave of
  intel sgx.
\newblock In {\em Proceedings of the Twenty-Fifth International Conference on
  Architectural Support for Programming Languages and Operating Systems},
  ASPLOS ’20, page 955–970, New York, NY, USA, 2020. Association for
  Computing Machinery.

\bibitem{ratel}
Shweta Shinde, Jinhua Cui, Satyaki Sen, Pinghai Yuan, and Prateek Saxena.
\newblock {Binary Compatibility For SGX Enclaves}, 2020.

\bibitem{shinde2015podarch}
Shweta Shinde, Shruti Tople, Deepak Kathayat, and Prateek Saxena.
\newblock Podarch: Protecting legacy applications with a purely hardware tcb.
\newblock {\em National University of Singapore, Tech. Rep}, 2015.

\bibitem{splash}
The parsec benchmark suite.
\newblock \url{https://parsec.cs.princeton.edu/license.htm}, 2020.

\bibitem{aegis}
G.~Edward Suh, Dwaine~E. Clarke, Blaise Gassend, Marten van Dijk, and Srinivas
  Devadas.
\newblock {AEGIS:} architecture for tamper-evident and tamper-resistant
  processing.
\newblock In {\em {ICS}}, pages 160--171. {ACM}, 2003.

\bibitem{tdx}
Intel® trust domain extensions.
\newblock
  \url{https://software.intel.com/content/www/us/en/develop/articles/intel-trust-domain-extensions.html}.

\bibitem{privado}
Shruti Tople, Karan Grover, Shweta Shinde, Ranjita Bhagwan, and Ramachandran
  Ramjee.
\newblock Privado: Practical and secure {DNN} inference.
\newblock {\em CoRR}, abs/1810.00602, 2018.

\bibitem{torch}
Torch | scientific computing for luajit.
\newblock \url{http://torch.ch}, 2020.

\bibitem{graphene-sgx}
Chia{-}che Tsai, Donald~E. Porter, and Mona Vij.
\newblock Graphene-sgx: {A} practical library {OS} for unmodified applications
  on {SGX}.
\newblock In {\em {USENIX} Annual Technical Conference}, 2017.

\bibitem{sfi}
Robert Wahbe, Steven Lucco, Thomas~E. Anderson, and Susan~L. Graham.
\newblock {Efficient Software-based Fault Isolation}.
\newblock In {\em Proceedings of the 14th ACM Symposium on Operating Systems
  Principles}, 1993.

\bibitem{xing2016sgx}
Bin~Cedric Xing, Mark Shanahan, and Rebekah Leslie-Hurd.
\newblock Intel® software guard extensions (intel® {SGX}) software support
  for dynamic memory allocation inside an enclave.
\newblock In {\em Proceedings of the Hardware and Architectural Support for
  Security and Privacy 2016}, HASP 2016, New York, NY, USA, 2016. Association
  for Computing Machinery.

\bibitem{sp3}
Jisoo Yang and Kang~G. Shin.
\newblock {Using Hypervisor to Provide Data Secrecy for User Applications on a
  Per-page Basis}.
\newblock In {\em VEE}, 2008.

\bibitem{nacl}
B.~{Yee}, D.~{Sehr}, G.~{Dardyk}, J.~B. {Chen}, R.~{Muth}, T.~{Ormandy},
  S.~{Okasaka}, N.~{Narula}, and N.~{Fullagar}.
\newblock Native client: A sandbox for portable, untrusted x86 native code.
\newblock In {\em 2009 30th IEEE Symposium on Security and Privacy}, 2009.

\bibitem{cloudvisor}
Fengzhe Zhang, Jin Chen, Haibo Chen, and Binyu Zang.
\newblock {CloudVisor: Retrofitting Protection of Virtual Machines in
  Multi-Tenant Cloud with Nested Virtualization}.
\newblock In {\em SOSP}, 2011.

\end{thebibliography}

\appendix

\begin{table*}[!tb]
\centering
\resizebox{\linewidth}{!}{\begin{tabular}{rlclcl}
\toprule
\multicolumn{2}{c}{\textbf{Interface}} & \textbf{Pre-condition} &
\multicolumn{3}{c}{\textbf{Transition Relation}} \\ \midrule
\texttt{create}
& $\begin{aligned}[t]
(p: \mPrincipal, \\ l: \mSize) \to \\ (r: \mRegion, e:
\mError) \end{aligned}$ & $p \in
\mP \land \exists{n: \mRegion} [n \notin \dom(\mR)]$ &
$\begin{aligned}[t]
\mSp = \mS[\mR / \mR + \{n \mapsto \langle p, l
\rangle\}, \\
\mA / \mA + \{\langle n, p \rangle \mapsto \langle \mathtt{rwxl},
\mathtt{rwxl} \rangle\}, \\
\mM / \mM + \{\langle n, i \rangle \mapsto 0 \mid i = 0, 1
\cdots, l - 1\}]
\end{aligned}$ &
$\land$ & 
$\begin{aligned}[t]
    r = n \\ \land \\ e = \mSuccess
\end{aligned}$\\
\midrule
\texttt{map} &
$\begin{aligned}[t]
(p: \mPrincipal, \\
v: \mVaddr,
    r: \mRegion) \to \\
    (e: \mError)
\end{aligned}$ &
$\begin{aligned}[t]
\langle r, p \rangle \in \dom(\mA) \land
    \forall{u: \mVaddr, g:
    \mRegion}\\ [\langle p, u, g\rangle  \notin \mV
    \lor \neg\mintersect\langle u, g, v, r\rangle ]
\end{aligned}$ &
$\begin{aligned}[t]
    \mSp = \mS[\mV/\mV + \{\langle p, v,
    r\rangle\}]
\end{aligned}$ &
$\land$ &
$e = \mSuccess$ 
\\ \midrule
\texttt{unmap} &
$\begin{aligned}[t]
(p: \mPrincipal, \\
    v: \mVaddr,
    r: \mRegion) \to \\
    (e: \mError)
\end{aligned}$ &
$\begin{aligned}[t]
    \langle p, v, r \rangle \in \mV
\end{aligned}$ &
$\mSp = \mS [\mV / \mV - \{\langle p, v, r \rangle\}]$ & 
$\land$ &
$e = \mSuccess$
\\ \midrule
\texttt{share} & $\begin{aligned}[t]
(p: \mPrincipal, \\
r: \mRegion,
o: \mPrincipal, \\
a: \mPermission) \to \\
(e: \mError) \end{aligned}$ &
$\begin{aligned}[t]
o \in \mP \land o \neq p \land p = \mR_{\text{Owner}}(r) \land \\
\langle r, o \rangle \notin \dom(\mA)
\end{aligned}$ &
$\begin{aligned}[t]
\mSp = \mS [\mA / \mA + \{\langle r, o \rangle \mapsto \langle
    a, \mbox{\texttt{-{}-{}-{}-}} \rangle\}
]
\end{aligned}$ & $\land$ & $e = \mSuccess$
\\ \midrule
\texttt{change} &
$\begin{aligned}[t]
(p: \mPrincipal, \\
r: \mRegion, \\
a: \mPermission) \to  \\
(e: \mError)
\end{aligned}$  &
$\begin{aligned}[t]
\langle r, p \rangle \in \dom(\mA) \land
a \leq \mA_{\text{MaxPerm}}(r, p) \land \\
(\mathtt{l} \in a \implies \forall{o \in \mP, o \neq a,
\langle r, o \rangle \in \dom(\mA)} \\
    [\mathtt{l} \notin \mA_{\text{Perm}}(r,
o)])
\end{aligned}$ & 
$\begin{aligned}[t]
\mSp = \mS[\mA / \mA[\langle r, p \rangle \mapsto \langle
\mA_{\text{MaxPerm}}(r, p), a \rangle]]
\end{aligned}$ & $\land$ & 
$e = \mSuccess$
\\ \midrule
\texttt{destroy} & 
$\begin{aligned}[t]
(p: \mPrincipal, \\ r: \mRegion) \to \\
(e: \mError)
\end{aligned}$ & 
$\begin{aligned}[t]
p = \mR_{\text{Owner}}(r)
\end{aligned}$ &
$\begin{aligned}[t]
\mSp = \mS[\mR/\mR - \{r \mapsto \langle p, l \rangle \mid l: \mSize
\}, \\
\mA/\mA - \{\langle r, o \rangle \mapsto \langle a, b \rangle \mid o:
\mPrincipal, \\ a: \mPermission, b: \mPermission\}, \\
\mM/\mM - \{\langle r, i \rangle \mapsto
    d \mid i: \mOffset, d: \mByte\},\\
    \mV/\mV - \{\langle p, v, r \rangle \mid v:
    \mVaddr\}]
\end{aligned}$ & $\land$ & $e = \mSuccess$
\\ \midrule
\texttt{transfer} & 
$\begin{aligned}[t]
(p: \mPrincipal, \\
r: \mRegion, \\
o: \mPrincipal) \to \\
(e: \mError)
\end{aligned}$ & 
$\begin{aligned}[t]
\langle r, p\rangle, \langle r, o\rangle \in \dom(\mA) \land
    \mathtt{l} \in \mA_{\text{Perm}}(r, p) \land \\
    \mathtt{l} \in \mA_{\text{MaxPerm}}(r, o) \\
\end{aligned}$ &
$\begin{aligned}[t]
\mSp = \mS[\mA / \mA[\langle r, p \rangle \mapsto \\ \langle
\mA_{\text{MaxPerm}}(r, p), \mA_{\text{Perm}}(r,
p) - \{\mathtt{l}\}\rangle, \\
\langle r, o \rangle \mapsto \\ \langle \mA_{\text{MaxPerm}}(r, o),
\mA_{\text{Perm}}(r,
o) + \{\mathtt{l}\} \rangle]] 
\end{aligned}$ & $\land$ & $e = \mSuccess$
\\ \midrule
\texttt{read} &
$\begin{aligned}[t]
(p: \mPrincipal, \\
    v: \mVaddr) \to \\
(d: \mByte, \\ e: \mError)
\end{aligned}$ & 
$\begin{aligned}[t]
    \exists{u:\mVaddr,
    r:\mRegion}[\langle p, u, r \rangle
    \in \mV \land \\
    \mcovers(u, r, v) \land
\mathtt{r} \in \mA_{\text{Perm}}(r, p) \land \\
    (\mathtt{l} \in \mA_{\text{Perm}}(r, p) \lor 
\forall{o \in \mP, \langle r, o \rangle \in \dom(\mA)}\\ [
\mathtt{l} \notin \mA_{\text{Perm}}(r, o)] )]
\end{aligned}$ & & & 
$\begin{aligned}[t]
d = \mM(r, v - u) \\ \land \\ e = \mSuccess
\end{aligned}$\\
\bottomrule
\end{tabular}
}
\caption{\codename Interface. 
\codename state is defined as: $\mS
    \coloneqq \langle \mP, \mR, \mA,
    \mM, \mV \rangle$, where
$\mP \coloneqq \{p \mid p: \mPrincipal\}$;
$\mR \coloneqq \mRegion \hookrightarrow \mPrincipal \times \mSize$;
$\mA \coloneqq \mRegion \times
    \mPrincipal \hookrightarrow
    \mPermission^2$;
$\mM \coloneqq \mRegion \times
    \mOffset \hookrightarrow \mByte$;
    and $\mV \subseteq \mPrincipal
    \times \mRegion \times \mVaddr$.
    $\dom(\cdot)$ denotes the domain of
    a function. 
$A \hookrightarrow B$ defines a partial
    function from set $A$ to set $B$
    (with a domain that is a subset of
    $A$). $\mR_{\text{Owner}}(r)$
    and $\mR_{\text{Size}}(r)$
    denote the owner the size (in
    bytes) of region $r$.
    $\mA_{\text{Perm}}(r, p)$ and
    $\mA_{\text{MaxPerm}}(r, p)$ to denote the
    dynamic permission and static
    maximum permission of region $r$
    with respect to enclave $p$.
    $\mR(r) \coloneqq \langle
    \mR_{\text{Owner}}(r),
    \mR_{\text{Size}}(r) \rangle,
    \mA(r, p) \coloneqq \langle
    \mA_{\text{MaxPerm}}(r, p),
    \mA_{\text{Perm}}(r, p) \rangle$.
    $\mPermission$ is defined as
    the power set of $\{\mathtt{r}, \mathtt{w},
    \mathtt{x}, \mathtt{l}\}$.
    $a \in \mPermission$ is represented as 
    $d_\mathtt{r}d_\mathtt{w}d_\mathtt{x}d_\mathtt{l}$,
    where each $d_x$ is either $x$,
    meaning that the permission bit $x$
    is present ($x \in a$), or
    $\mathtt{-}$, meaning $x \notin a$. In the
    transition relations,
    $A[B/B^\prime]$ means replacing $B$
    with $B^\prime$ in $A$ and keeping
    everything else the same; $A[a
    \mapsto b]$ means changing the value
    of $A(a)$ to $b$ while keeping
    everything else the same.
    $\mintersect(u, g, v, r)$ is defined
    as $u + \mR_{\text{Size}}(g) > v
    \land v + \mR_{\text{Size}}(r) > u$.
    $\mcovers(u, r, v)$ is defined as $u
    \leq v < u + \mR_{\text{Size}}(r)$.
    Apart from explicit instructions parameters listed in
    Table~\ref{tab:instructions}, 
    we defined an
    extra argument $p: \mPrincipal$,
    to represent the enclave which invokes the
    instruction. We define
    a memory region \texttt{read}.
    We omit \texttt{write} and
    \texttt{execute}; their
    behavior is similar to
    \texttt{read}.
    }
    \label{tab:interface_def}
\end{table*}
 
\section{\codename Interface}
\label{sec:appx}
We present the formal definitions for each of the instruction in 
\codename. Table~\ref{tab:interface_def} gives the detailed pre-condition
checks and the transition relations for each instruction.

\end{document}